\documentclass[twocolumn]{aastex61}

\received{4th April 2017}
\revised{}
\accepted{}
\submitjournal{ApJ}

\shorttitle{SAMI cluster kinematic-morphology relation}
\shortauthors{Brough et al.}

\begin{document}

\title{The SAMI Galaxy Survey: mass as the driver of the kinematic morphology -- density relation in clusters}

\author{Sarah Brough}
\affil{School of Physics, University of New South Wales, NSW 2052, Australia}
\affil{Australian Astronomical Observatory, 105 Delhi Rd, North Ryde, NSW 2113, Australia}
\affil{ARC Centre of Excellence for All-sky Astrophysics (CAASTRO)}

\author{Jesse van de Sande}
\affiliation{Sydney Institute for Astronomy (SIfA), School of Physics, The University of Sydney, NSW 2006, Australia}

\author{Matt S. Owers}
\affil{Australian Astronomical Observatory, 105 Delhi Rd, North Ryde, NSW 2113, Australia}
\affiliation{Department of Physics and Astronomy, Macquarie University, NSW 2109, Australia}

\author{Francesco d'Eugenio}
\affil{ARC Centre of Excellence for All-sky Astrophysics (CAASTRO)}
\affiliation{Research School of Astronomy and Astrophysics, Australian National University, Canberra, ACT 2611, Australia}

\author{Rob Sharp}
\affiliation{Research School of Astronomy and Astrophysics, Australian National University, Canberra, ACT 2611, Australia}

\author{Luca Cortese}
\affiliation{International Centre for Radio Astronomy Research (ICRAR), University of Western Australia, 35 Stirling Highway, Crawley, WA 6009, Australia}

\author{Nicholas Scott}
\affil{ARC Centre of Excellence for All-sky Astrophysics (CAASTRO)}
\affiliation{Sydney Institute for Astronomy (SIfA), School of Physics, The University of Sydney, NSW 2006, Australia}

\author{Scott M. Croom}
\affil{ARC Centre of Excellence for All-sky Astrophysics (CAASTRO)}
\affiliation{Sydney Institute for Astronomy (SIfA), School of Physics, The University of Sydney, NSW 2006, Australia}

\author{Rob Bassett}
\affiliation{International Centre for Radio Astronomy Research (ICRAR), University of Western Australia, 35 Stirling Highway, Crawley, WA 6009, Australia}

\author{Kenji Bekki}
\affiliation{International Centre for Radio Astronomy Research (ICRAR), University of Western Australia, 35 Stirling Highway, Crawley, WA 6009, Australia}

\author{Joss Bland-Hawthorn}
\affiliation{Sydney Institute for Astronomy (SIfA), School of Physics, The University of Sydney, NSW 2006, Australia}

\author{Julia J. Bryant}
\affil{Australian Astronomical Observatory, 105 Delhi Rd, North Ryde, NSW 2113, Australia}
\affil{ARC Centre of Excellence for All-sky Astrophysics (CAASTRO)}
\affiliation{Sydney Institute for Astronomy (SIfA), School of Physics, The University of Sydney, NSW 2006, Australia}

\author{Roger Davies}
\affiliation{Astrophysics, Department of Physics, University of Oxford, Denys Wilkinson Building, Keble Rd., Oxford, OX1 3RH, UK}

\author{Michael J. Drinkwater}
\affil{ARC Centre of Excellence for All-sky Astrophysics (CAASTRO)}
\affil{School of Mathematics and Physics, University of Queensland, QLD 4072, Australia}

\author{Simon P. Driver}
\affiliation{International Centre for Radio Astronomy Research (ICRAR), University of Western Australia, 35 Stirling Highway, Crawley, WA 6009, Australia}

\author{Caroline Foster}
\affil{Australian Astronomical Observatory, 105 Delhi Rd, North Ryde, NSW 2113, Australia}

\author{Gregory Goldstein}
\affiliation{Department of Physics and Astronomy, Macquarie University, NSW 2109, Australia}

\author{\'A. R. L\'opez-S\'anchez}
\affil{Australian Astronomical Observatory, 105 Delhi Rd, North Ryde, NSW 2113, Australia}
\affiliation{Department of Physics and Astronomy, Macquarie University, NSW 2109, Australia}

\author{Anne M. Medling}
\affiliation{Research School of Astronomy and Astrophysics, Australian National University, Canberra, ACT 2611, Australia}
\affiliation{Cahill Center for Astronomy and Astrophysics, California Institute of Technology, MS 249-17 Pasadena, CA 91125, USA}
\affiliation{Hubble Fellow}

\author{Sarah M. Sweet}
\affiliation{Centre for Astrophysics and Supercomputing, Swinburne University of Technology, PO Box 218, Hawthorn, VIC 3122, Australia}

\author{Dan S. Taranu}
\affil{ARC Centre of Excellence for All-sky Astrophysics (CAASTRO)}
\affiliation{International Centre for Radio Astronomy Research (ICRAR), University of Western Australia, 35 Stirling Highway, Crawley, WA 6009, Australia}

\author{Chiara Tonini}
\affiliation{School of Physics, Melbourne University, Parkville, VIC 3010 Australia}

\author{Sukyoung K. Yi}
\affiliation{Department of Astronomy and Yonsei University Observatory, Yonsei University, Seoul 120-749, Republic of Korea}

\author{Michael Goodwin}
\affil{Australian Astronomical Observatory, 105 Delhi Rd, North Ryde, NSW 2113, Australia}

\author{J. S. Lawrence}
\affil{Australian Astronomical Observatory, 105 Delhi Rd, North Ryde, NSW 2113, Australia}

\author{Samuel N. Richards}
\affil{Australian Astronomical Observatory, 105 Delhi Rd, North Ryde, NSW 2113, Australia}
\affil{ARC Centre of Excellence for All-sky Astrophysics (CAASTRO)}
\affiliation{Sydney Institute for Astronomy (SIfA), School of Physics, The University of Sydney, NSW 2006, Australia}



\begin{abstract}

We examine the kinematic morphology of early-type galaxies (ETGs) in eight galaxy clusters in the Sydney-AAO Multi-object Integral field spectrograph (SAMI) Galaxy Survey.  The clusters cover a mass range of $14.2<\log(M_{200}/M_{\odot})<15.2$ and we measure spatially-resolved stellar kinematics for 315 member galaxies with stellar masses $10.0<\log(M_*/M_{\odot})\leq11.7$ within $1R_{200}$ of the cluster centers.  We calculate the spin parameter, $\lambda_R$ and use that to classify the kinematic morphology of the galaxies as fast or slow rotators. The total fraction of slow rotators in the early-type galaxy population, $F_{SR}=0.14\pm{0.02}$ and does not depend on host cluster mass. Across the eight clusters, the fraction of slow rotators increases with increasing local overdensity.  We also find that the slow-rotator fraction increases at small clustercentric radii ($R_{cl}<0.3R_{200}$), and note that there is also an increase in slow-rotator fraction at $R_{cl}\sim0.6R_{200}$. The slow rotators at these larger radii reside in cluster substructure. We find the strongest increase in slow-rotator fraction occurs with increasing stellar mass. After accounting for the strong correlation with stellar mass, we find no significant relationship between spin parameter and local overdensity in the cluster environment.  We conclude that the primary driver for the kinematic morphology--density relationship in galaxy clusters is the changing distribution of galaxy stellar mass with local environment.  The presence of slow rotators in substructure suggests that the cluster kinematic morphology--density relationship is a result of mass segregation of slow-rotating galaxies forming in groups that later merge with clusters and sink to the cluster center via dynamical friction.

\end{abstract}


\section{Introduction}

The relative fraction of different galaxy morphological types has been shown to vary with environment such that galaxies visually classified as early-type galaxies are more prevalent in the high-density environment of galaxy clusters at the expense of late-type galaxies.  This is the morphology--density relationship \citep{oemler74, davis76, dressler80}.  Early-type galaxies (ETGs) can also be classified based on a kinematic morphology using their spin parameter, probed through their stellar kinematics.  In this classification system, early-type galaxies with high spin parameter are classified as fast rotators (FRs) and those with low spin parameter are classified as slow rotators (SRs; \citealt{cappellari07,emsellem07}; see \citealt{cappellari16} for a review).

The ATLAS$^{\rm{3D}}$ team examined the kinematic morphology--density relationship for the first time \citep{cappellari11_2}.  
They observed that there are few slow-rotating early-type galaxies, relative to the total number of early-type galaxies, in the lowest density local environments such that the fraction of slow-rotating ETGs, $F_{SR}=N_{SR}/N_{ETG}\sim0.13$. However, the fraction of slow rotators more than doubles in the densest region of the Virgo cluster where $F_{SR}\sim0.28$.  
Virgo is not a very massive cluster and provides only a single example of a dense environment.  The kinematic morphology--density relationship has since been studied in seven additional clusters of differing cluster masses (Abell 1689, \citealt{deugenio13}; Coma, \citealt{houghton13}; Fornax, \citealt{scott14} and Abell 85, 168 and 2399, \citealt{fogarty14}). These authors all find a total slow rotator fraction, $F_{SR}\sim0.15$ with no dependence on the global environment studied between the field/group sample of ATLAS$^{\rm{3D}}$ and the most massive cluster studied to-date, Abell 1689.  However, these studies also find that $F_{SR}$ generally rises with increasing local environmental density within those global environments.  The Abell 168 and Abell 2399 clusters studied by \cite{fogarty14} form an exception. In these two clusters the slow rotator fraction peaks at intermediate densities and then falls.  These two clusters are known to be undergoing mergers \citep{hallman04, fogarty14} and the slow rotators in these systems are associated with cluster substructure.  
\cite{houghton13} argued that the relationship observed between kinematic morphology and density is a result of mass segregation by dynamical friction because the total slow rotator fraction is consistent across a range of global environments, while the slow rotators are segregated into the densest local environments. \cite{cappellari16} came to the same conclusion owing to the presence of slow rotators near the centers of the Fornax, Virgo and Coma clusters or subgroups within those clusters.

There is also a known relationship between galaxy mass and spin parameter such that the highest mass, more luminous galaxies show the lowest spin parameters (e.g. \citealt{emsellem07, jimmy13, cappellari13, veale16, oliva-altamirano16}). While low-mass dwarf galaxies have a strong relationship between spin parameter and environment \citep{toloba15,guerou15} it is not yet clear whether the driving force in the kinematic morphology--density relationship for the general elliptical galaxy population is environmental density or galaxy mass.  \cite{scott14} examined this question for a sample of 30 early-type galaxies in the Fornax cluster and found that even in mass-matched samples of slow and fast rotators, the slow rotators were found at preferentially higher projected environmental density than the fast rotators. They argued that dynamical friction alone, therefore, could not be responsible for the differing distributions of slow and fast rotators.

The Sydney-AAO Multi-object Integral field spectrograph (SAMI; \citealt{croom12}) now makes it possible to obtain spatially-resolved optical spectra for large numbers of galaxies covering a broad range in mass and environment.  The SAMI Galaxy Survey \citep{bryant15} will observe $\sim3600$ galaxies with stellar masses $7<\log(M_*/M_{\odot})<12$ in a range of environments including eight galaxy clusters \citep{owers17}.  We present here the kinematic morphology--density relationship for the early-type galaxies observed in the eight SAMI clusters.  This is the largest sample of cluster galaxies available to-date and allows a robust analysis of the dependence of kinematic morphology on stellar mass as well as global and local environment. 

In Section~\ref{sect:observations} we describe our observations and data reduction.  Derived parameters are described in Section~\ref{sect:derived} and our kinematic classification is defined in Section~\ref{sect:kinematics}. Results are presented in Section~\ref{sect:results}, and discussed in Section~\ref{sect:discussion} before conclusions are drawn in Section~\ref{sect:conclusions}. Throughout this paper we assume a Hubble constant of $H_0=70$ km s$^{-1}$ Mpc$^{-1}$, $\Omega_M=0.3$, $\Omega_\Lambda=0.7$ and a \cite{chabrier03} initial mass function.

\section{Observations and Sample}
\label{sect:observations}
The observations used in this analysis are selected from the SAMI Galaxy Survey.  The survey is ongoing and will observe $\sim3600$ galaxies at redshifts $0.04<z<0.095$ with regular public data releases which will be accessible via the survey website, https://sami-survey.org (Green et al., in prep).  
The survey target selection is described in \cite{bryant15}. In brief, the main SAMI sample is selected from the Galaxy And Mass Assembly survey (GAMA; \citealt{driver11,hopkins13}).  
The GAMA sample covers broad ranges in stellar mass and environment but does not include massive clusters. The SAMI Galaxy Survey also targets eight additional clusters \citep{owers17} to probe higher density environments.  

Following previous analyses of the kinematic morphology--density relation, we focus here on the SAMI cluster sample; a separate paper will analyse the kinematic morphology--density relation in the main SAMI sample (van de Sande et al., in prep). The selection of the eight clusters and their constituent galaxies is described in detail in \cite{owers17}. In brief, cluster members were selected following a dedicated redshift program using the AAOmega spectrograph \citep{sharp06} on the 3.9m Anglo-Australian Telescope (AAT) fed by its multi-object fiber-feed: 2dF.  The redshift survey has high spectroscopic completeness and 94 per cent of potential cluster members 
have redshift measurements. Cluster membership was defined using a caustic analysis \citep{owers17}.  The cluster members were used to estimate $R_{200}$\footnote{The radius at which the mean interior density is 200 times the critical density of the Universe.}, as well as the mass within $R_{200}$, $M_{200}$.  The SAMI clusters range from $14.2<\log(M_{200}/M_{\odot})<15.2$ in mass and $0.02<z<0.06$ in redshift (Table~\ref{tab:clusters}). 

\begin{figure}
	\includegraphics[width=22pc]{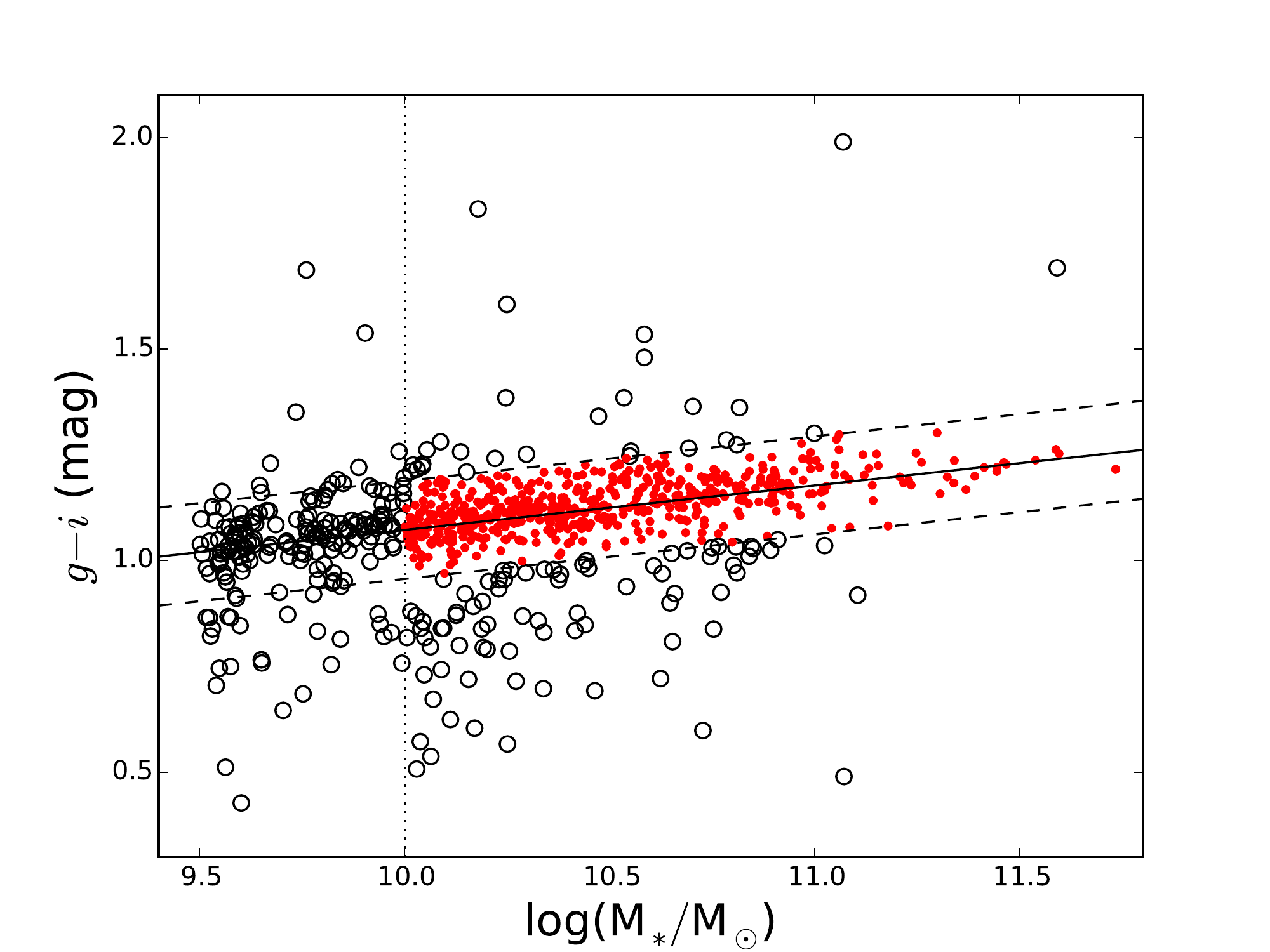}
    \caption{$g-i$ color as a function of stellar mass, $M_*$, for the 848 SAMI cluster members (within 1R$_{200}$ and $\pm3.5V_{gal}/\sigma_{cl}$ and stellar mass $\log M_*/M_{\odot}>9.5$ for $z_{cl}<0.045$ and $\log M_*/M_{\odot}>10.0$ for $z_{cl}>0.045$; black open points) and those galaxies selected for this analysis (red galaxies with $\log M_*/M_{\odot}\geq10.0$; red filled points).  The solid line indicates a straight-line fit to the red sequence and the dashed line indicates the $\pm1\sigma$ scatter around that. The dotted line indicates the stellar mass limit $\log M_*/M_{\odot}=10.0$.}
    \label{fig:color_mass}
\end{figure}

SAMI has a stepped stellar mass selection function so if the cluster redshift is less than $z=0.045$ the stellar mass limit is $\log M_*/M_{\odot}=9.5 $ and clusters with redshifts above that have a stellar mass limit of  $\log M_*/M_{\odot}=10.0 $.  SAMI targets cluster members within $1R_{200}$ and $\pm3.5V_{gal}/\sigma_{cl}$ (the cluster-centric recession velocity with respect to the cluster velocity dispersion) and there are 848 cluster members meeting these stellar mass, radius and recession velocity criteria. The stacked color-stellar mass distribution of the 848 cluster members is illustrated in Figure~\ref{fig:color_mass}.

Previous analyses of the kinematic morphology--density relation focussed on the early-type galaxies. We do not have visual morphologies for all of the galaxies in our sample. The majority of visually-classified early-type galaxies are red (although not all, e.g. \citealt{bassett17}), particularly in the cluster environment. We therefore make a color cut to select only the red galaxies (within $\pm1\sigma$ of the fitted red sequence, equivalent to $\pm0.11~$mag) in the $g-i$ color-mass relationship for the stacked clusters (c.f. \citealt{houghton13}). This selection is illustrated in Figure~\ref{fig:color_mass}. The precise choice of color cut does not change our conclusions.

As lower mass galaxies were not observed by SAMI above $z=0.045$, to maximise sample completeness over the whole redshift range we only select galaxies with stellar masses $\log M_*/M_{\odot}>10.0$ for the analysis presented here. The mass and color cuts remove 181 and 108 galaxies respectively, leaving a sample of 559 early-type cluster members with $\log M_*/M_{\odot}>10.0$ (Table~\ref{tab:clusters}). Henceforth, we do not include cluster members with $\log M_*/M_{\odot}<10.0$ in our analysis.



\begin{table*}
	\centering
	\caption{The properties of the eight galaxy clusters observed by SAMI. R.A., Dec, z$_{cl}$, M$_{200}$, R$_{200}$ and $\sigma_{cl}$ are all from Owers et al. (2017). N$_{Mem}$ is the number of members within 1R$_{200}$ and $\pm3.5V_{gal}/\sigma_{cl}$ (for clusters with $z_{cl}<0.045$ this includes galaxies with $9.5<\log M_*/M_{\odot}<10.0$). N$_{ETG}$ is the number of early-type members with $\log M_*/M_{\odot}\geq10.0$ (ETGs). N$_{Obs}$ is the number of observed ETGs. N$_{\lambda}$ is the number of observed ETGs for which the spin parameter could be measured. Comp$_{ETG,\lambda}$ gives the fraction of ETGs with spin parameter measurements. N$_{SPS}$ gives the number of SAMI Pilot Survey galaxies added to the analysis (Section~4.2.2).}
	\label{tab:clusters}
\begin{tabular}{ccccccccccccc}
\hline
Cluster & R.A.& Dec & z$_{cl}$ & M$_{200}$ & R$_{200}$ & $\sigma_{cl}$ & N$_{Mem}$ & N$_{ETG}$ & N$_{Obs}$ & N$_{\lambda}$ & Comp$_{ETG,\lambda}$ & N$_{SPS}$ \\
&J2000&J2000& &$\log(M_{\odot}$)&Mpc& kms$^{-1}$& &  &  &  & &  \\
\hline
  EDCC0442 & 6.38068 & -33.04657 & 0.0498 & 14.45 & 1.41 & 583 & 50 & 42 & 33 & 31 & 0.74 & 0\\
  Abell0085 & 10.460211 & -9.303184 & 0.0549 & 15.19 & 2.42 & 1002& 167 & 138 & 65 & 58 & 0.42 & 12\\
  Abell0119 & 14.06715 & -1.25537 & 0.0442 & 14.92 & 2.02 & 840 & 253 & 138 & 64 & 55 & 0.40 & 0\\
  Abell0168 & 18.815777 & 0.213486 & 0.0449 & 14.28 & 1.33 & 546 & 112 & 53 & 21 & 21 & 0.40 & 8\\
  Abell2399 & 329.389487 & -7.794236 & 0.0579 & 14.66 & 1.63 & 690 & 92 & 73 & 54 & 49 & 0.67 & 3\\
  Abell3880 & 336.97705 & -30.575371 & 0.0578 & 14.64 & 1.62 & 660 & 56 & 48 & 28 & 28 & 0.58 & 0\\
  APMCC0917 & 355.39788 & -29.236351 & 0.0509 & 14.26 & 1.19 & 492 & 29 & 23 & 18 & 15 & 0.65 & 0\\
  Abell4038 & 356.93781 & -28.140661 & 0.0293 & 14.36 & 1.46 & 597 & 89 & 44 & 37 & 35 & 0.80 & 0\\
\hline
  Total: & - & - & - & - & - & - & 848 & 559 & 320 & 292 & 0.52 & 23\\
      \hline\end{tabular}

\end{table*}


\subsection{SAMI Observations and Data Reduction}

The SAMI instrument \citep{croom12} deploys 13 imaging fiber bundles, `hexabundles' (\citealt{bland-hawthorn11}; \citealt{bryant14}), over a 1 degree field at the Prime Focus of the AAT. Each hexabundle consists of 61 circularly packed optical fibers. The core size of each fiber is 1.6 arcsec, giving each hexabundle a field-of-view of 15 arcsec diameter. All 819 fibers (793 object fibers and 26 sky fibers) feed into the AAOmega spectrograph. For SAMI observing, AAOmega is configured to a wavelength coverage of 370--570 nm with R = 1812 in the blue arm, and 630--740 nm with R = 4263 in the red arm \citep{vandesande16}. A seven point dither pattern achieves near-uniform spatial coverage \citep{sharp15}, with 1800 s exposure time for each frame, totalling 3.5 h per field.

As described in \cite{allen15}, in every field, 12 galaxies and a secondary standard star are observed. The secondary standard star is used to probe the conditions as observed by the entire instrument. The flux zero-point is obtained from SDSS while the shape of the flux correction is derived from primary standard stars observed in a single hexabundle during the same night for any given field of observation. The raw data from SAMI were reduced using the AAOmega data reduction pipeline, 2dfDRv5.62 \citep{croom04, sharp10} followed by full alignment and flux calibration through the SAMI Data Reduction pipeline (see \citealt{sharp15} for a detailed explanation of this package). In addition to the reduction pipeline described by \cite{allen15} and \cite{sharp15}, the individual frames are now scaled to account for variations in observing conditions (Green et al., in prep). 

To-date the SAMI Galaxy Survey has observed 320 early-type cluster member galaxies with $\log M_*/M_{\odot}\geq10$ (SAMI internal data release v0.9.1).


The 8 clusters include 3 clusters (Abell 85, 168 and 2399) observed previously as part of the SAMI Pilot Survey \citep{fogarty14, fogarty15}.  Some of these galaxies have been re-observed and the data reduction improved. We present a comparison with the analysis here in Section~\ref{ssect:spinparam}.

\section{Derived Parameters}
\label{sect:derived}

\subsection{Photometry}

The photometry for the SAMI Galaxy Survey clusters is described in detail in \cite{owers17}. We provide a brief overview here. 

Four of the clusters (Abell 85, 119, 168 and 2399) lie within the Sloan Digital Sky Survey (SDSS; \citealt{york00}). For these clusters the $ugriz$ SDSS DR10 photometry has been re-measured using the IOTA software used to measure aperture-matched photometry for the GAMA survey \citep{hill11,driver16}. Each frame was convolved to a common point spread function (PSF) full-width at half maxiumum (FWHM)$=2^{\prime\prime}$ before using the SExtractor software \citep{bertin96} in dual-image mode to extract the aperture- and seeing-matched photometry.  The $r$-band image was used for detection.

The four clusters not in the SDSS regions (EDCC 0442, APMCC0917, Abell 3880 and 4038) are covered by the Very large telescope Survey Telescope (VST) ATLAS \citep{shanks15} survey.  Raw VST/ATLAS data in the $gri$-bands were retrieved from the archive and reduced using the Astro-WISE optical image reduction pipeline \citep{mcfarland13}. The aperture- and PSF-matched photometry is measured as for the SDSS data, with one exception: the image quality of the VST/ATLAS imaging is higher than that of SDSS (Shanks et al. 2015), so each 1 deg~$\times$~1 deg $gri$-band tile was convolved to a common $1.5^{\prime\prime}$ FWHM. 
\cite{owers17} use the duplicate measurements of galaxies in Abell 85, which has full SDSS and partial VST/ATLAS coverage, to show that any systematic differences in the photometric and stellar mass measurements between the two surveys are less than 0.05 dex.

To calculate absolute magnitudes, we K-correct the apparent magnitudes to $z = 0$ using the IDL calc\_kor.pro code \citep{chilingarian12}\footnote{http://kcor.sai.msu.ru}.

\subsection{Stellar Masses}

Stellar masses are estimated using the empirical proxy between $i$-band absolute magnitude and $g-i$ color given in \cite{taylor11} and also used by \cite{bryant15} for the main SAMI Galaxy Survey.  We use the aperture- and PSF-matched photometry described above, corrected for Galactic extinction using \cite{schlegel98} dust maps.

\subsection{Photometric Fits}

Galaxy effective radii and ellipticities are measured using the Multi Gaussian Expansion (MGE; \citealt{emsellem94}) technique implemented in the code from \cite{cappellari02}. These are presented in d'Eugenio et al. (in prep.).  The code measures circularized effective radii which we convert to the semi-major axis effective radius used throughout this paper as, $R_e=R_{e,circ} / (\sqrt{1-\epsilon})$, where the ellipticity, $\epsilon$, is the luminosity-weighted ellipticity of the galaxy.


\subsection{Galaxy Environment}
\label{ssect:environment}

The local environment of galaxies can be measured using a nearest-neighbor surface density to probe the underlying density field. The principle behind the nearest-neighbor measurement is that galaxies with closer neighbours are in denser environments \citep{muldrew12}.  The nearest-neighbor measurement can be refined to an overdensity which parametrises whether galaxies are in an environment more or less dense than the average in a given sample.  We calculate the nearest-neighbor surface density, $\Sigma_{N,Vlim,Mlim}$, for all galaxies with reliable redshifts in the parent cluster redshift sample. The surface density is defined using the projected co-moving distance to the \mbox{Nth-nearest} neighbor ($d_N$) with a velocity limit $\pm V_{lim}$ km s$^{-1}$: $\Sigma_{N,Vlim,Mlim}=N/\pi d_N$.  The neighbors are all within a volume-limited density-defining population that has absolute magnitudes $M_r < M_{lim} - Qz$. $Q$ defines the expected evolution of $M_r$ as a function of redshift, $z$, ($Q = 1.03$; \citealt{loveday15}).  The large spatial extent of the SAMI Cluster Redshift Survey (extending to $2R_{200}$; \citealt{owers17}) means that when measuring nearest-neighbor surface densities within $1R_{200}$ no Nth-nearest neighbors are separated from the galaxy in question by more than the distance to the 2R200 cluster `edge'.  Such large separations would rapidly increase the uncertainty of such measurements.
We also measured the overdensity, $\delta_{N,Vlim,Mlim}=\Sigma / \bar{\Sigma}$,  dividing the density by the mean density of the early-type members with $\log M_*/M_{\odot}>10$.  In Appendix~\ref{appendix:densities} we analyse the effect the choice of limits has on the nearest-neighbor surface density and overdensity. We find that the nearest-neighbor surface density is sensitive to the choice of limits applied, while the overdensity is not. We therefore choose an overdensity of $\delta_{5,500,-18.3}$ in this analysis.


\section{Kinematic Classification}
\label{sect:kinematics}

\subsection{Stellar kinematics}
\label{ssect:stellarkin}
The stellar kinematic measures made for the SAMI survey are described in detail in \cite{vandesande16}. We summarise the salient points here.  

The mean line-of-sight stellar velocity, $V$, and velocity dispersion, $\sigma$ are measured using the penalized pixel fitting code (pPXF; \citealt{cappellari04}).  The SAMI Galaxy Survey runs pPXF in two different modes.  All results presented here consist of fits using a Gaussian line of sight velocity dispersion (LOSVD).  
To measure the stellar kinematics, the spectra from the blue and red arms of the spectrograph are combined. Before this can happen the red spectra (FWHM$_{red}=1.61~\AA$) are convolved to the instrumental resolution of the blue spectra (FWHM$_{blue}=2.65~\AA$).  

Optimal templates are constructed for 1-5 annular bins per galaxy (depending on signal-to-noise ratio; S/N) by running pPXF over the combined spectra using the full MILES stellar library \citep{sanchezblazquez06}.  After this, pPXF is run three times with only the optimal templates for each galaxy spaxel.  The first run is used to estimate the real noise from the residuals.  The second run uses the new noise spectrum for masking emission lines and bad pixels.  The third run derives the LOSVD parameters.  For each spaxel pPXF is allowed to use the optimal templates from the annular bin in which the spaxel lives as well as from neighboring annuli. The uncertainties on the LOSVD are standard deviations after fitting pPXF to 150 simulated spectra.  To construct the simulated spectra the best-fit template is first subtracted from the spectrum. The residuals, together with the noise spectrum, are then randomly rearranged in wavelength space within eight wavelength sectors. The residuals, are added to the best-fit template to construct the simulated spectra, which are then refitted with pPXF.  We compared the measurement uncertainties we obtain from these 150 simulations with the pPXF uncertainty estimates and find that they agree well \citep{vandesande16}. However, looking at the 2D uncertainty maps, the simulated spectra provide less stochastic uncertainty maps that are more consistent with the S/N of the galaxy spectra so we use these in our analysis.

We apply a quality cut that ensures we keep a large fraction of the low velocity dispersion spaxels, while keeping a strict quality cut for the higher velocity dispersions \citep{vandesande16}: Only spaxels with velocity dispersions $\sigma>$ FWHM$_{\rm{instr}}/2\sim35$ km s$^{-1}$, and velocity and velocity dispersion uncertainties, $\Delta V < 30$km s$^{-1}$ and \mbox{$\Delta \sigma < (\sigma*0.1) + 25$km s$^{-1}$} are retained in the final analysis. These quality cuts result in a sample with a median velocity dispersion uncertainty at S/N$~<20~\AA^{-1}$ is 12.6 per cent and 2.6 per cent for S/N$~>20~\AA^{-1}$ (Figure~2, \citealt{vandesande16}).

  
\subsection{Spin parameter}
\label{ssect:spinparam}
\begin{figure}
	\includegraphics[width=22pc]{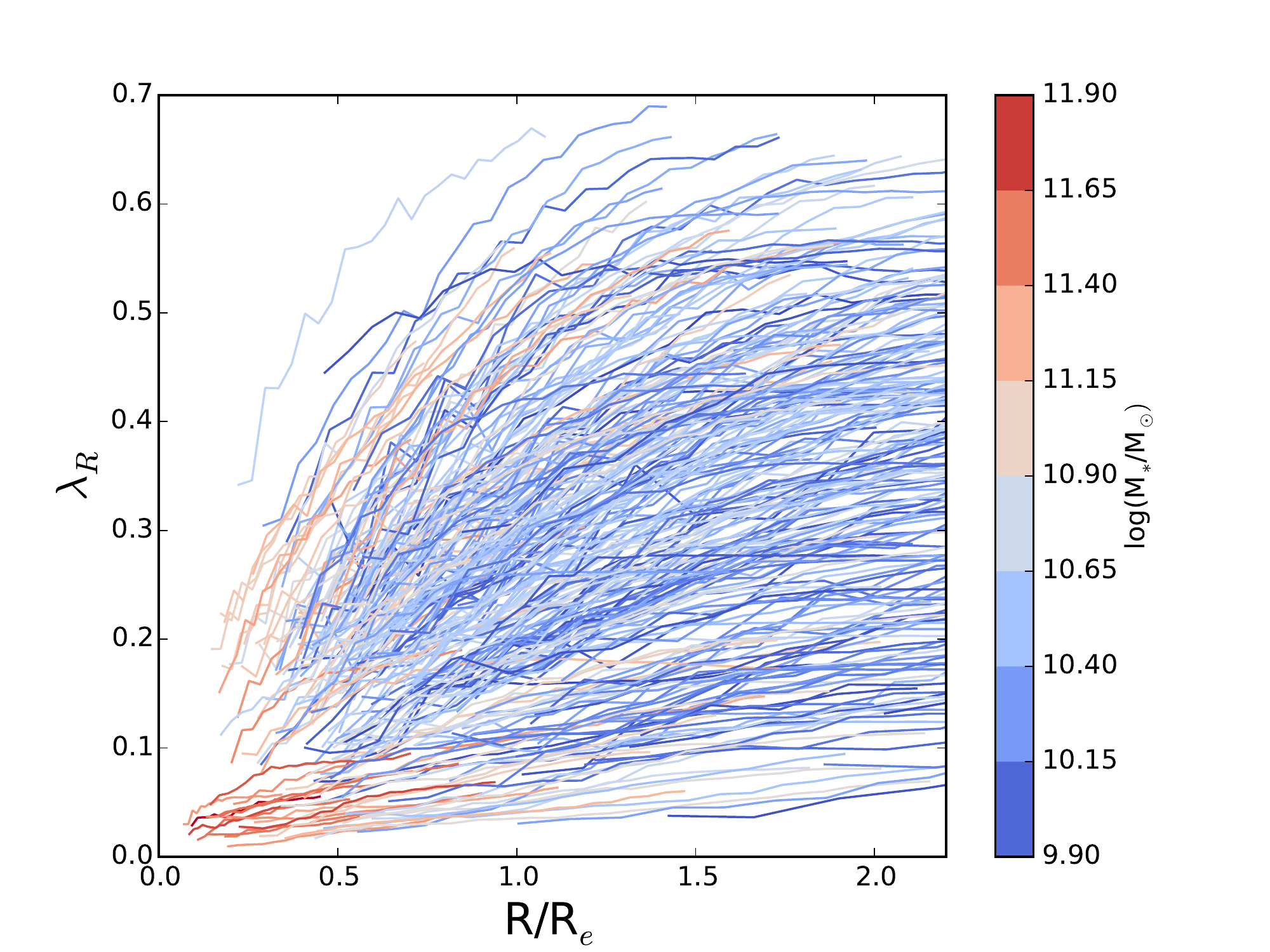}
    \caption{Spin parameter profiles, $\lambda_R$, as a function of normalised galaxy radius, $R/R_e$.  Points within the seeing radius ($R_{PSF}\sim1.5^{\prime\prime}$) are not plotted. The colors indicate stellar mass, $M_*$.}
    \label{fig:lambda_prof}
\end{figure}

\cite{emsellem07} defined the luminosity-weighted spin parameter ($\lambda_{R}$):

\begin{equation}
\centering
\label{eq:lambda}
\lambda_{R}= \frac{\sum^{i=N}_{i=0}F_iR_i|V_i|}{\sum^{i=N}_{i=0}F_iR_i\sqrt{V_i^2+\sigma_i^2}}  ,
\end{equation}
in this analysis $R_i$ is the semi-major radius of the ellipse in which spaxel $i$ is located and $F_i$ is the flux of the $i^{th}$ spaxel. 
$\lambda_R$ is summed over all spaxels, $N$, that meet the quality cut described above within the ellipse of semi-major axis $R$.  The $\lambda_R$ profiles are illustrated in Figure~\ref{fig:lambda_prof}.

The spin parameter is summed within a fiducial radius, $R_{fid}$ which can be either $0.5R_e$, $1R_e$ or $2R_e$ (following previous analyses).  
A particular fiducial radius is only used when $R_{fid}>R_{PSF}\sim1.5^{\prime\prime}$ and the percentage of spaxels within that radius that meet the quality cut is $>75$ per cent.  Our first choice is to measure $\lambda_{R_{fid}}$ within $1R_e$. However, if the effective radius is smaller than $R_{PSF}$ we use $R_{fid}=2 R_e$ and if the galaxy has $R_e>15^{\prime\prime}$ then we use $R_{fid}=0.5R_e$.  224 galaxies have $R_{fid}=1R_e$, 46 have $R_{fid}=2R_e$, and 19 have $R_{fid}={0.5R_e}$. Three of the exceptionally large brightest cluster galaxies in these clusters have semi-major effective radii $>15^{\prime\prime}$ and so the fiducial radius for their $\lambda$ measurements are $\sim0.3R_e$.  We cannot measure $\lambda$ at all for 28 galaxies due to: nearby galaxies affecting their observation (N~$=22$) and too low S/N observations (N~$=6$).  This leaves a sample of 292 galaxies for which we can measure $\lambda_{R_{fid}}$.

\begin{figure*}
\center
	\includegraphics[width=30pc]{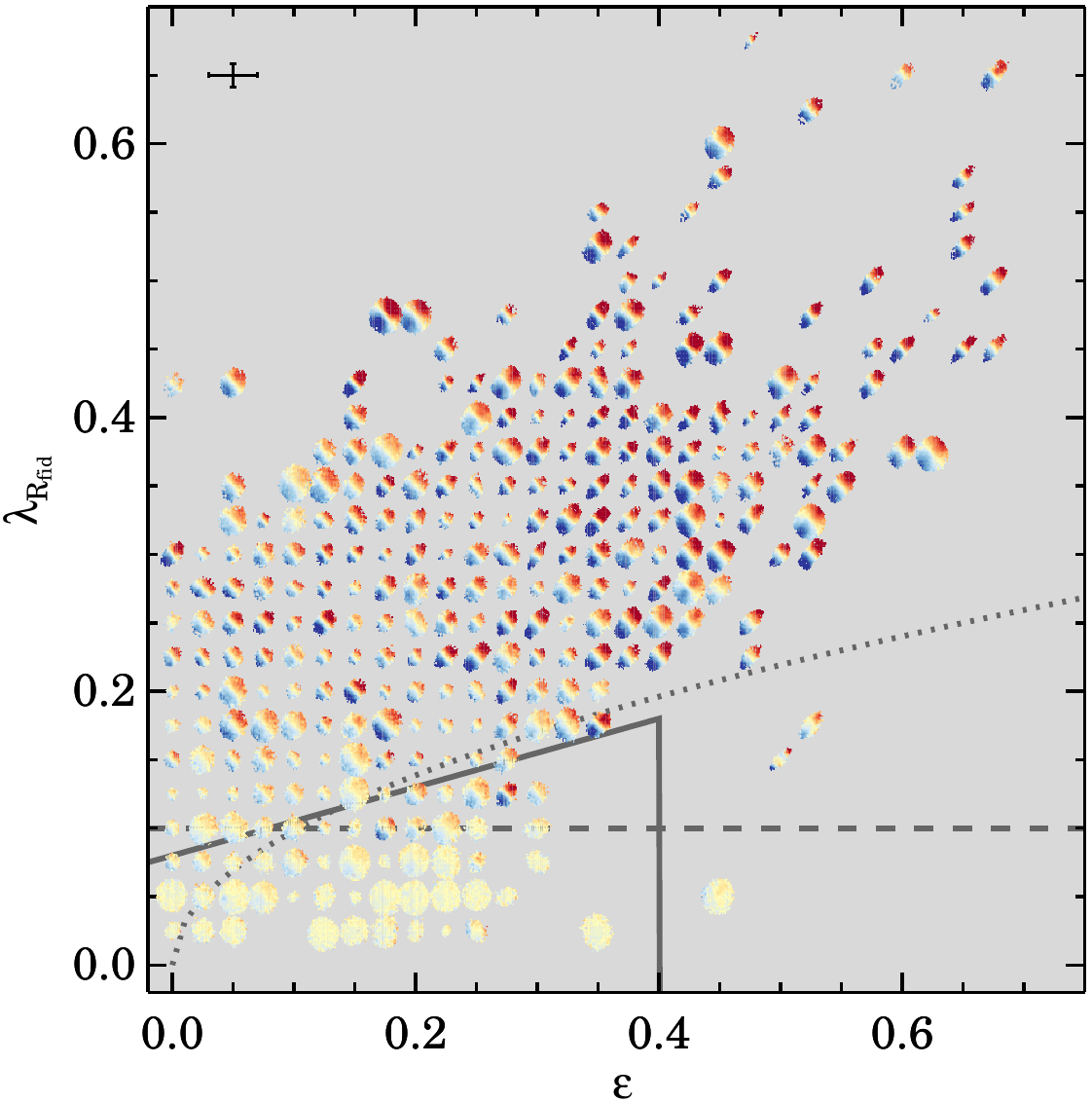}
    \caption{Spin parameter, $\lambda_{R_{fid}}$, as a function of ellipticity, $\epsilon$.  The lines indicate the Cappellari et al. (2016; solid), Emsellem et al. (2011; dotted) and Emsellem et al. (2007; dashed) fast/slow rotator separations. The average measurement uncertainty is shown in the top left-hand corner. For each galaxy we show its stellar velocity map aligned to $45^{\circ}$ using the kinematic position angle, with the scale of the velocity color map set by the stellar mass Tully-Fisher relation. A grid is applied to avoid overlap of the velocity maps. }
    \label{fig:lambda_eps}
\end{figure*}

The spin parameter as a function of ellipticity is shown in Figure~\ref{fig:lambda_eps}.  This plot has been constructed in the same way as the equivalent in \cite{vandesande16}. The figures are not identical as here we focus on the cluster galaxies and do not study the higher-order stellar kinematics so we have a lower S/N cut and include galaxies with $\lambda$ measured within fiducial radii other than $1R_e$. For each galaxy in Figure~\ref{fig:lambda_eps}, we show the velocity map to highlight the stellar velocities. To avoid overlap between the galaxy velocity maps, the data are first put on a regular grid with a spacing of 0.02 in $\lambda_{R_{fid}}$ and $\epsilon$. We position each galaxy on its closest grid point, or its neighbor if its closest grid point is already filled by another galaxy. The size of the grid and velocity maps are chosen such that no galaxy is offset by more than one grid point from its original position. The stellar mass -- Tully-Fisher \citep{dutton11} relation is used for the velocity map color scale: for a galaxy with stellar mass $\log M_*/M_{\odot}>10$ the scale of the velocity map ranges from $-95 < V ($km s$^{-1}) < 95$, whereas a galaxy with stellar mass $\log M_*/M_{\odot}>11$ is assigned a velocity range from $-169 < V ($km s$^{-1}) < 169$. The kinematic position angle is used to align the major axis of all galaxies to $45^{\circ}$. The velocity maps are truncated where the S/N is too low ($<3$) and the errors do not meet the quality criteria. This truncation is different for every galaxy.

\subsubsection{Choice of Fiducial Radius}
Measuring $\lambda_{R}$ within a fiducial radius of $0.5R_{e}$, or less, could potentially affect our findings. It might introduce a bias as it is generally the more massive galaxies for which we are unable to reach a fiducial radius of $1R_{e}$. We test the effect of this by taking the 224 galaxies with $\lambda$ measured at $1R_{e}$ and find the value of $\lambda$ at $0.5R_{e}$ for those galaxies.  Because $\lambda$ increases with increasing radius (Figure~\ref{fig:lambda_prof}), measuring $\lambda$ at radii less than $1R_{e}$ is likely to bias the measured $\lambda$ lower, artificially inflating the number of slow-rotating galaxies. Therefore, we focus on the 26/224 test galaxies that have $\lambda_{0.5R_e}\leq0.1$. The mean change in $\lambda_R$ measured at $0.5R_e$ compared to $1R_e$, \mbox{$\lambda_{0.5R_e}$/$\lambda_{1R_e}=0.58\pm0.03$}, with no correlation with stellar mass. We test the effect of that offset by applying it to the 19 galaxies for which we can only measure $\lambda_{R}$ within $0.5R_{e}$, or less.  Applying the offset does decrease the fraction of slow rotators but it 
does not change any of the conclusions we draw in the remainder of this paper.  We present the remainder of the results using the uncorrected $\lambda_{R_{fid}}$ values.




%

\subsubsection{Including SAMI Pilot Survey Observations}
\begin{figure}
	\includegraphics[width=22pc]{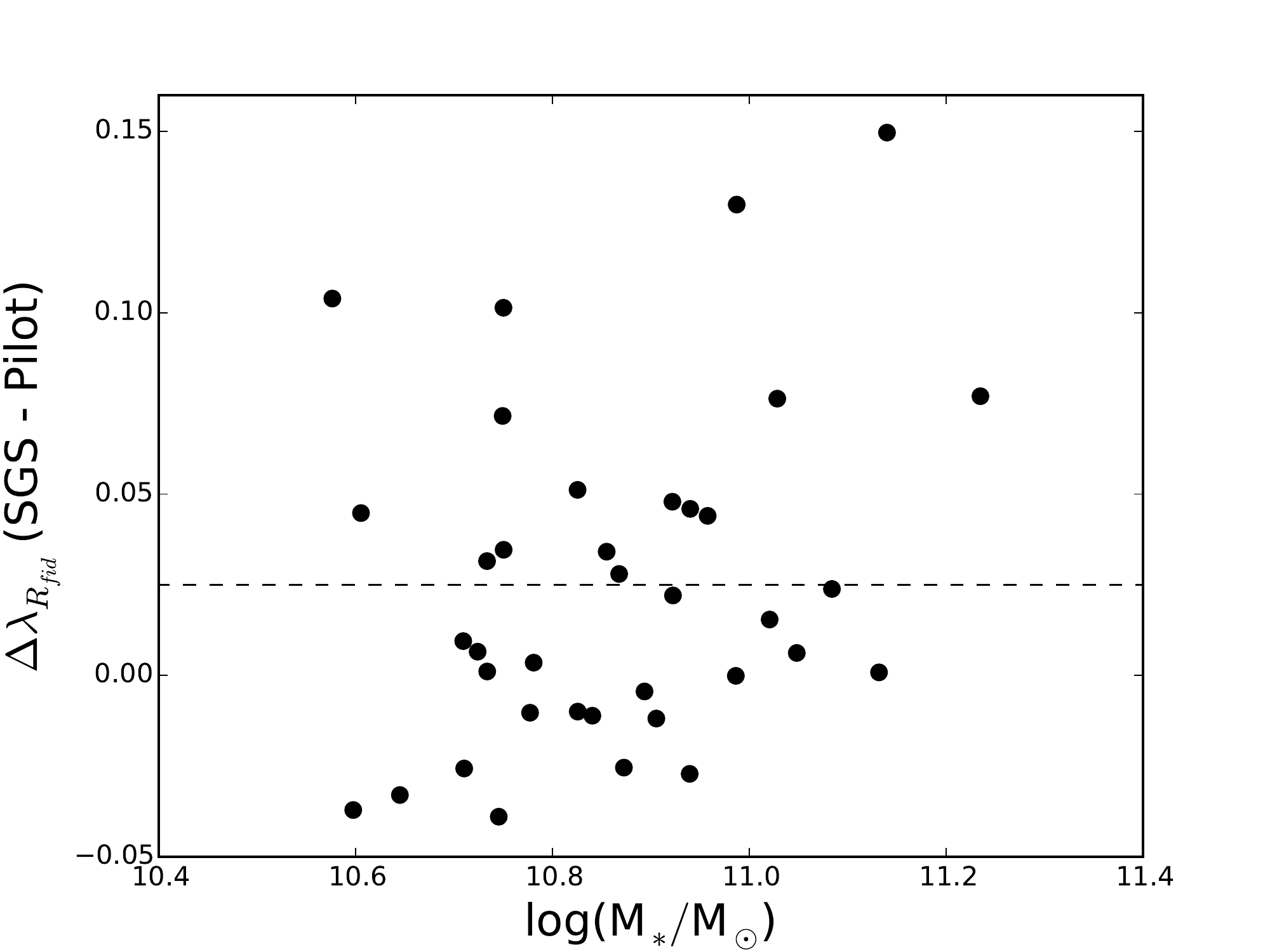}
    \caption{ Comparing spin parameter measurements for 37 galaxies in common between the SAMI Galaxy Survey (SGS) and SAMI Pilot Survey (Pilot) with their stellar masses. The mean difference, $\lambda_{R_{fid}}(SGS-Pilot)=0.025\pm0.007$, is shown by the dashed line. There is no significant offset in spin parameter between the two surveys.} 
    \label{fig:pilot_comp}
\end{figure}

The SAMI Pilot Survey \citep{fogarty14,fogarty15} includes stellar kinematics measured for 106 galaxies in 3 of the clusters (Abell 85, 168 and 2399) presented here.  The Pilot Survey stellar kinematics were calculated in the same way as described here (Section~\ref{ssect:stellarkin}) and their stellar masses and colors have been measured here.  Of the 106 pilot survey galaxies, 78 are within $1R_{200}$ of their cluster center and 69 of those 78 meet the color and stellar mass selection criteria we apply here.  Of the 69 that meet our criteria, 46 have been re-observed to-date as part of the SAMI Galaxy Survey (SGS). We use slightly different quality criteria here that mean that only 37/46 of these galaxies use the same fiducial radius to measure the spin parameter.  In Figure~\ref{fig:pilot_comp} we compare the Pilot Survey spin parameters with those measured here and find a mean difference $\lambda_{R_{fid}}(SGS-Pilot)=0.025\pm0.007$ for these 37 galaxies.   There are some outliers in this distribution. Our method for determining the stellar kinematics and sizes and ellipticities have been improved and when we compare the kinematic maps for the outliers we find holes due to the exclusion of low S/N data in the \cite{fogarty15} maps that are not present in our data. We note that the fast and slow rotator classifications do not change between the two surveys.  We therefore include the 23 Pilot Survey galaxies that meet our selection criteria but have not yet been re-observed as part of the SAMI Galaxy Survey in our analysis from this point.  The inclusion of these galaxies does not affect the conclusions we draw.  

Adding the 23 SAMI Pilot Survey galaxies to the 292 SAMI Galaxy Survey galaxies 
gives us a final sample of 315 galaxies. 

\subsection{Sample Completeness}
We are interested in fractional quantities, so it is important to understand the observed completeness of our sample.

\begin{figure}
	\includegraphics[width=22pc]{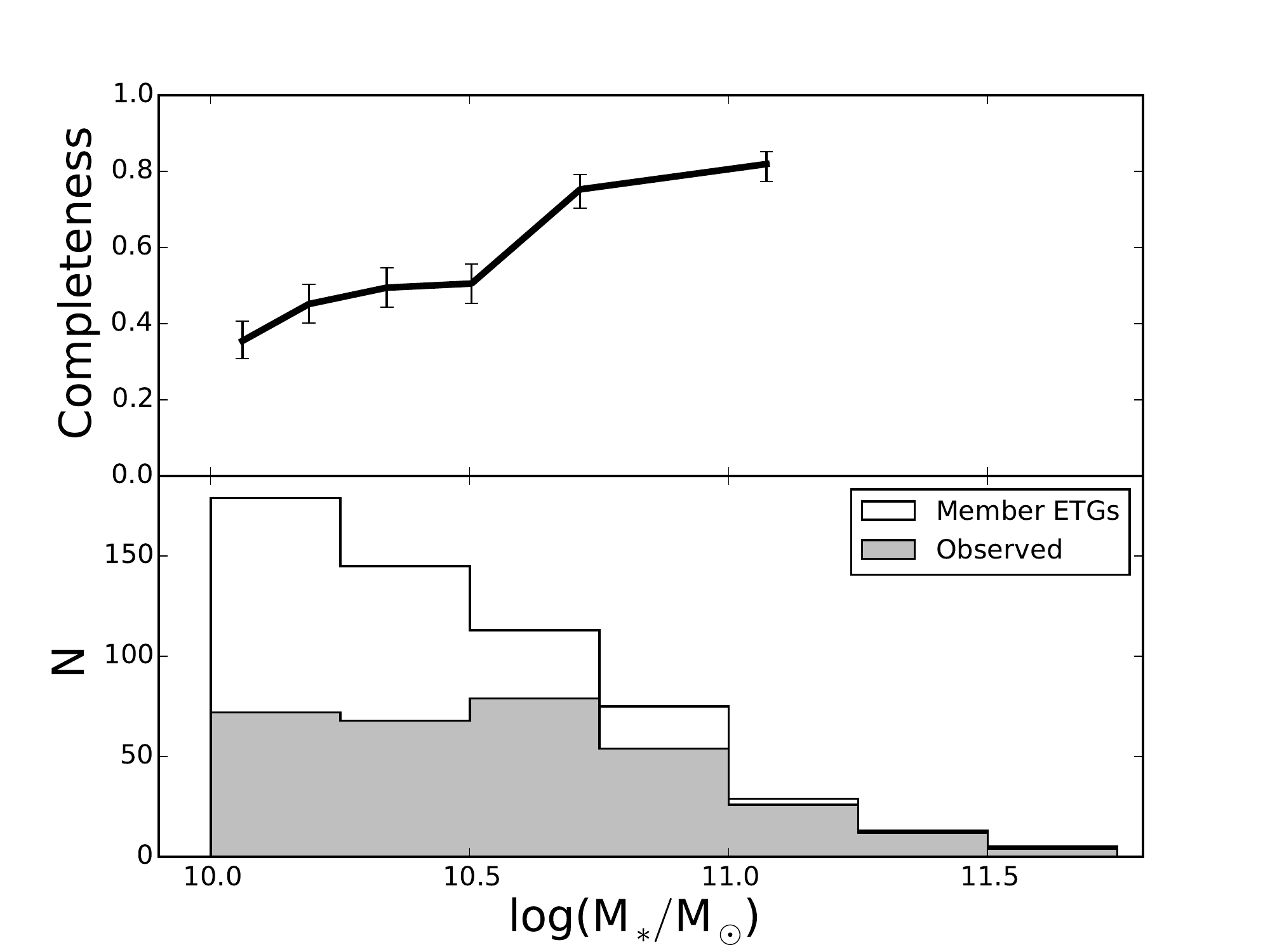}
    \caption{Completeness of galaxies with spin parameter measurements as a function of stellar mass. The lower panel shows the stellar mass distribution of early-type members with $\log M_*/M_{\odot}>10$ (Member ETGs), and those with spin parameter measurements (Observed) while the upper panel shows the fraction of these (Completeness) as a function of stellar mass. The error bars show the $1\sigma$ binomial confidence limits on these measurements. The observed completeness rises as a function of increasing stellar mass.}
    \label{fig:mass_comp}
\end{figure}

The completeness of galaxies with spin parameter measurements as a function of stellar mass is shown in Figure~\ref{fig:mass_comp}.  The observed completeness rises as a function of increasing stellar mass as a result of early targetting decisions \citep{owers17}. We analyse the effect of this further in Section~\ref{subsect:f_sr}.


\begin{figure}
	\includegraphics[width=22pc]{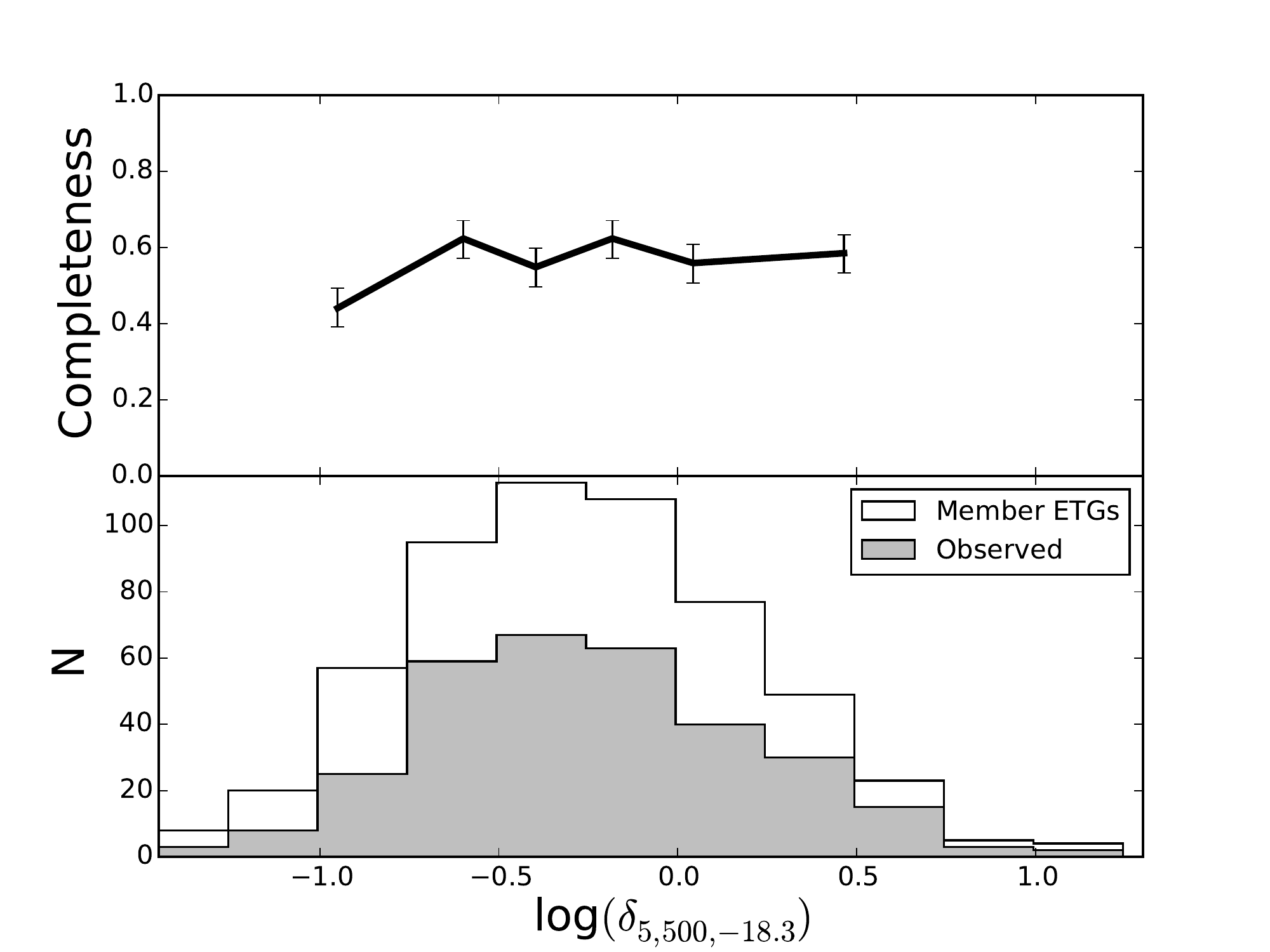}
    \caption{Completeness of galaxies with spin parameter measurements as a function of overdensity. The lower panel shows the overdensity distribution of early-type members with $\log M_*/M_{\odot}>10$ (Member ETGs), and those with spin parameter measurements (observed) while the upper panel shows the fraction of these (Completeness) as a function of  overdensity. The error bars show the $1\sigma$ binomial confidence limits on these measurements. Observed completeness is flat as a function of overdensity.}
    \label{fig:env_comp}
\end{figure}

The completeness of galaxies with spin parameter measurements as a function of overdensity is shown in Figure~\ref{fig:env_comp}.  We find that the observed completeness is flat as a function of environmental overdensity.

\subsection{Slow/Fast Rotator Separation}

Using a quantitative analysis of stellar velocity maps, the ATLAS$^{\rm{3D}}$ team \citep{krajnovic08,krajnovic11} found that their sample of 260 early-type galaxies broke into broad groupings of fast- and slow-rotating galaxies that could be separated using the following definitions from Emsellem et al. (2011; Equations~\ref{eq:emsellem1} and \ref{eq:emsellem2}) and Fogarty et al. (2014; Equation~\ref{eq:fogarty}): 
\begin{eqnarray}   
\label{eq:emsellem1} 
\lambda_{0.5R_e}<0.265\sqrt{\epsilon_{0.5R_e}}\\
\label{eq:emsellem2} 
\lambda_{R_e}<0.31\sqrt{\epsilon_{R_e}} \\
\label{eq:fogarty} 
\lambda_{2R_e}<0.363\sqrt{\epsilon_{2R_e}} 
\end{eqnarray}
The relationship for $1R_e$ is shown as the dotted line in Figure~\ref{fig:lambda_eps}. All kinematic morphology--density relationship analyses have used these definitions to separate their early-type galaxy samples into fast and slow rotators. Figure~\ref{fig:lambda_eps} also shows a new definition for separating fast- and slow-rotating galaxies from Cappellari (2016; solid line):
\begin{equation}
\lambda_{R_e}<0.08+\epsilon_e/4    ~\rm{with}~\epsilon_e<0.4
\label{eq:cappellari}
\end{equation}
as well as the classification from \citealt{emsellem07}:
\begin{equation}
\lambda_{R_e}<0.1
\label{eq:emsellem}
\end{equation}
shown as the dashed line in Figure~\ref{fig:lambda_eps}. Similar to the ATLAS$^{\rm{3D}}$ team, we see broad groups of fast- and slow-rotating galaxies in Fig~\ref{fig:lambda_eps}.  Using the \cite{emsellem11} definition there are 30 slow rotators in our sample, in comparison to 42 using the \cite{cappellari16} definition and 38 using the \cite{emsellem07} definition.  Inspecting Figure~\ref{fig:lambda_eps}, the \cite{emsellem11} slow rotator definition appears not to select some slow-rotating galaxies, while both the \cite{emsellem07} and \cite{cappellari16} definitions capture all the slow-rotating galaxies.  We, therefore, use the \cite{cappellari16} definition in the remainder of this paper but note that the choice of slow/fast rotator separation does not affect the conclusions we draw.  

\section{Results}
\label{sect:results}

\subsection{Fraction of Slow Rotators}
\label{subsect:f_sr}

We investigate here the fraction of slow-rotating galaxies as a function of the number of early-type galaxies, $F_{SR}$.  Throughout this section we plot $F_{SR}$ in bins of equal numbers of galaxies with fractional uncertainties calculated using binomial confidence intervals shown to be accurate for small to intermediate sample sizes \citep{cameron11}.

\begin{figure}
	\includegraphics[width=22pc]{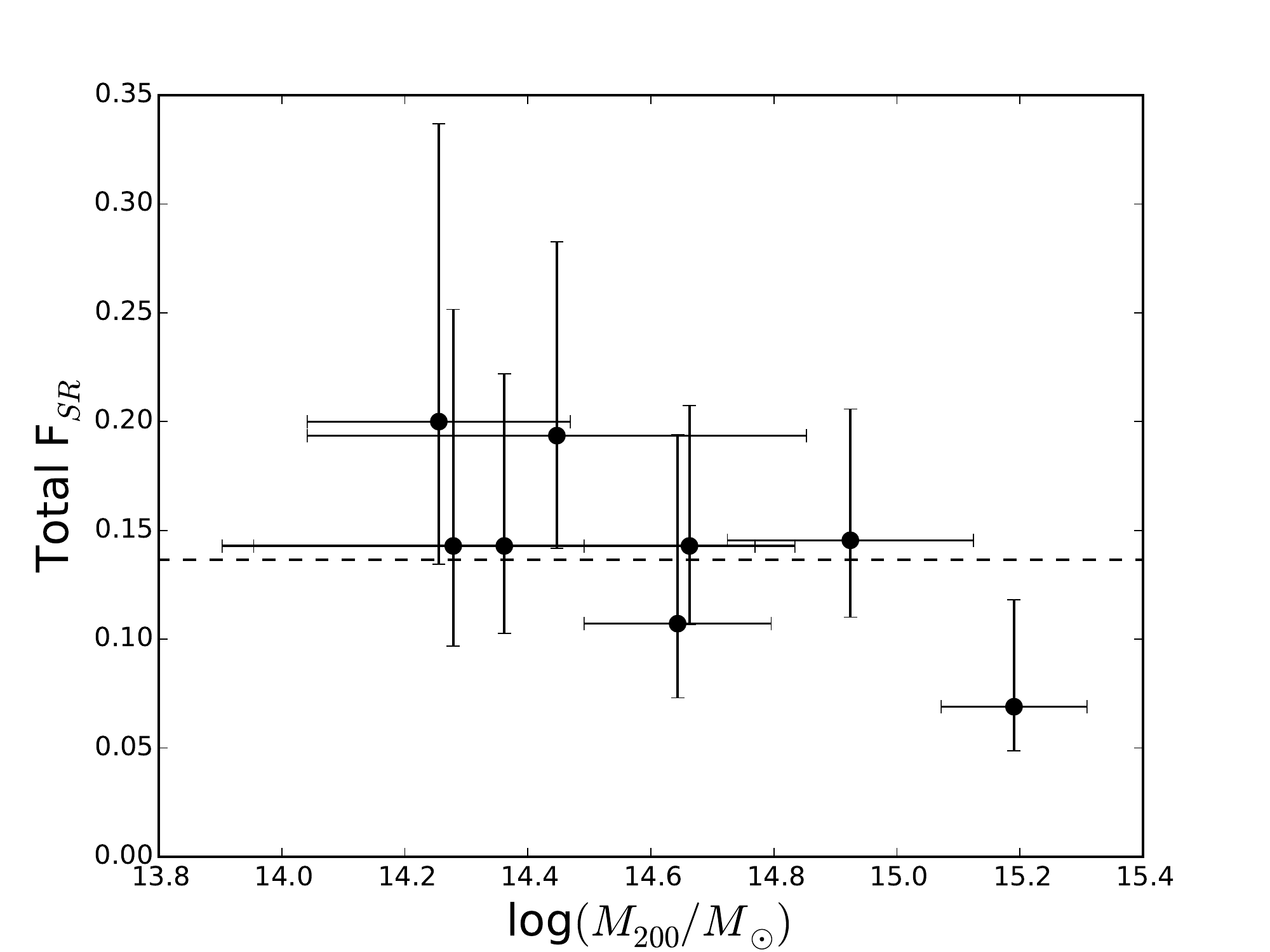}
	
    \caption{Total fraction of slow rotators, $F_{SR}$, as a function of host cluster mass, $M_{200}$. The dashed line shows the total fraction of slow rotators across our whole sample. 
    The uncertainties are the fractional uncertainties. We find no relationship of total $F_{SR}$ with host cluster mass over this mass range.}
    \label{fig:frac_clustermass}
\end{figure}

The total fraction of slow rotators in our sample is $F_{SR}=0.14\pm{0.02}$.  We examine the total fraction of slow rotators per cluster as a function of host cluster mass in Figure~\ref{fig:frac_clustermass} and find no significant relationship over the mass range examined here.

\begin{figure*}
	\includegraphics[width=45pc]{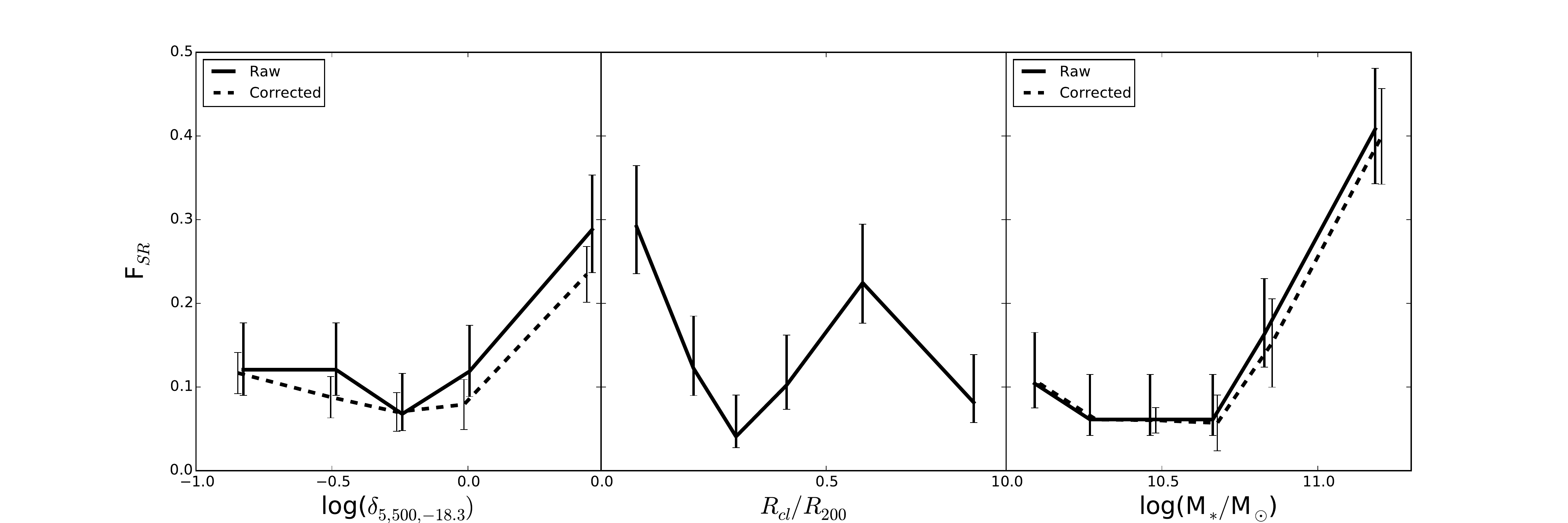}
    \caption{Fraction of slow rotators, $F_{SR}$. The left-hand panel shows $F_{SR}$ as a function of environmental overdensity, $\delta_{5,500,-18.3}$. The solid line gives the observed data with the heavy error bars showing the fractional uncertainties. The dashed line indicates the results of testing the effect of higher SAMI observing completeness at higher stellar masses. The light error bars are the standard deviation on the Monte Carlo analysis, offset in overdensity for visibility. Stellar mass completeness does not have a significant effect on the $F_{SR}$--$\delta_{5,500,-18.3}$ relation.  We observe an increasing fraction of slow rotators with increasing overdensity. The middle panel shows $F_{SR}$ as a function of stacked cluster-centric distance, $R_{cl}/R_{200}$. The fraction of slow rotators increases with decreasing cluster-centric radius.  Interestingly there is a `bump' at $R_{cl}/R_{200}\sim0.6$ due to substructure in four of the clusters. The right-hand panel shows $F_{SR}$ as a function of stellar mass, $M_*$.  The solid line gives the observed data with the heavy error bars showing the fractional uncertainties. The dashed line indicates the results of testing the effect of higher SAMI observing completeness at higher stellar masses. The light error bars are the standard deviation on the Monte Carlo analysis, offset in stellar mass for visibility. Stellar mass completeness does not have a significant effect on the $F_{SR}$--$M_*$ relation. We observe that the fraction of slow rotators increases with increasing stellar mass. }
    \label{fig:frac3}
\end{figure*}

In the left-hand panel of Figure~\ref{fig:frac3} we examine the fraction of slow-rotating galaxies, $F_{SR}$, as a function of local environmental overdensity.  
We calculate the significance of our results by comparing the fraction of slow rotators in the four lowest overdensity bins (which are statistically equal) to the fraction in the highest overdensity bin, taking into account the uncertainties in this measurement: Significance $=(F_{SR, high \delta} - F_{SR, low \delta})/\sqrt(\sigma_{F_{SR,high \delta}}^2+\sigma_{F_{SR,low\delta}}^2$). We find an increasing fraction of slow rotators with increasing overdensity with a significance of $3.4 \sigma$.

We note that there is a higher SAMI observing completeness at higher stellar masses due to early targetting decisions (Figure~\ref{fig:mass_comp}). This bias could affect the $F_{SR}-\delta$ relationship. We test this by Monte Carlo re-sampling the observed data. We determine the lowest completeness in stellar mass (40 per cent at $\log M_*/M_{\odot}\sim10$), and random re-sampling galaxies with stellar masses above $\log M_*/M_{\odot}\sim10.5$ down to that lowest observed completeness. We re-calculate the slow rotator fraction as a function of overdensity for each of 100 random re-samplings.  The mean $F_{SR,corrected}$ is shown by the dashed line in the left-hand panel of Figure~\ref{fig:frac3} and is indistinguishable from the observed relationship within the uncertainties.

\begin{figure}
	\includegraphics[width=22pc]{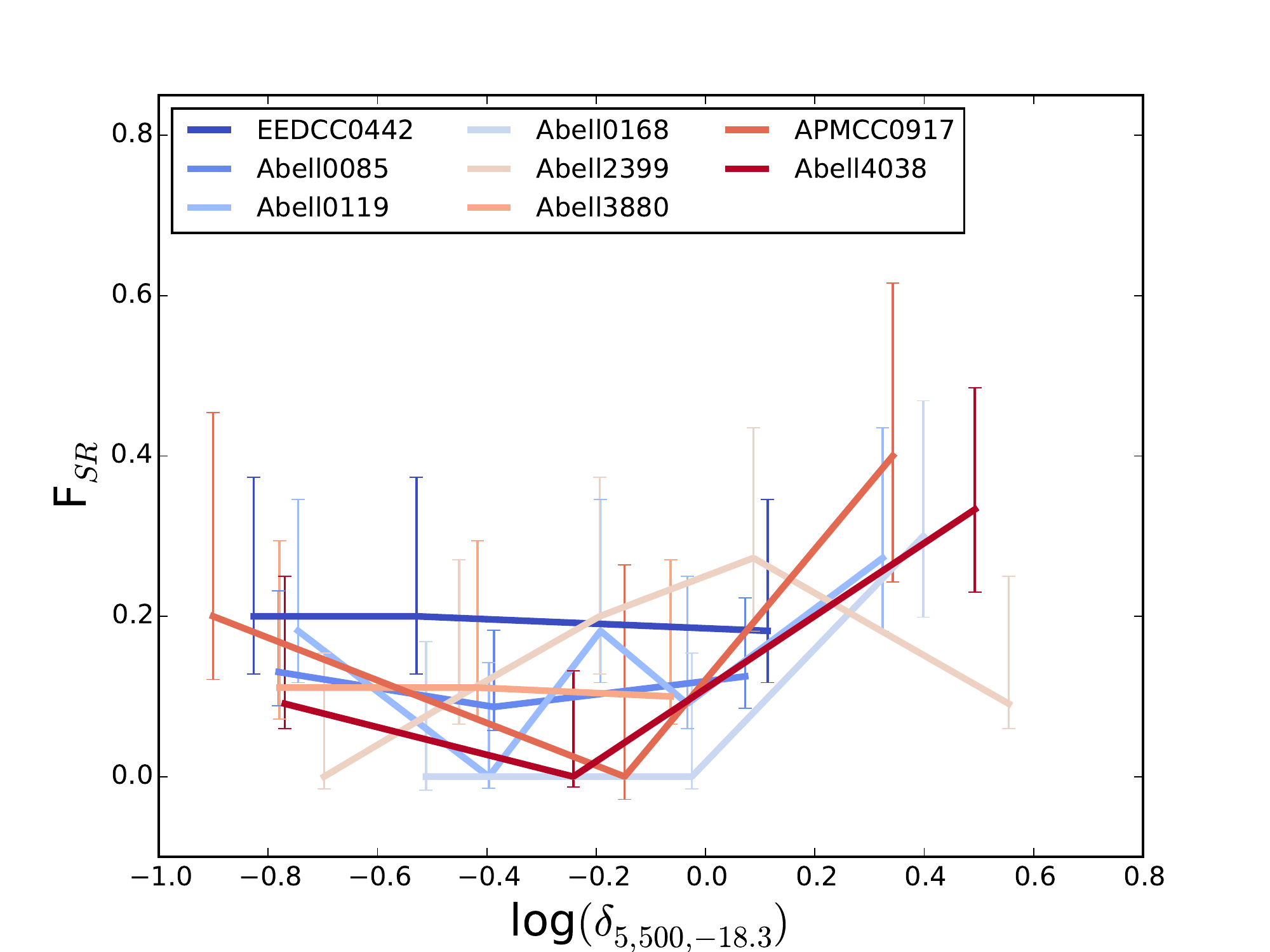}
    \caption{Fraction of slow rotators, $F_{SR}$, as a function of environmental overdensity, $\delta_{5,500,-18.3}$, for each individual cluster. The individual cluster fractions are noisier than the stacked distribution but show a general increase of $F_{SR}$ with overdensity.  }
    \label{fig:frac_env_indiv}
\end{figure}

In Figure~\ref{fig:frac_env_indiv} we examine $F_{SR}$ as a function of the environment overdensity for each of the 8 clusters in the SAMI survey.  These fractions are noisier than the stacked relationship shown in Figure~\ref{fig:frac3} and do not all rise with increasing overdensity.  
Like \cite{fogarty14}, we find that the slow rotator fraction within Abell 2399 peaks at intermediate overdensities and then drops, although we note that the uncertainties are large so the radial change is not statistically significant.

\begin{figure*}
\includegraphics[width=42pc]{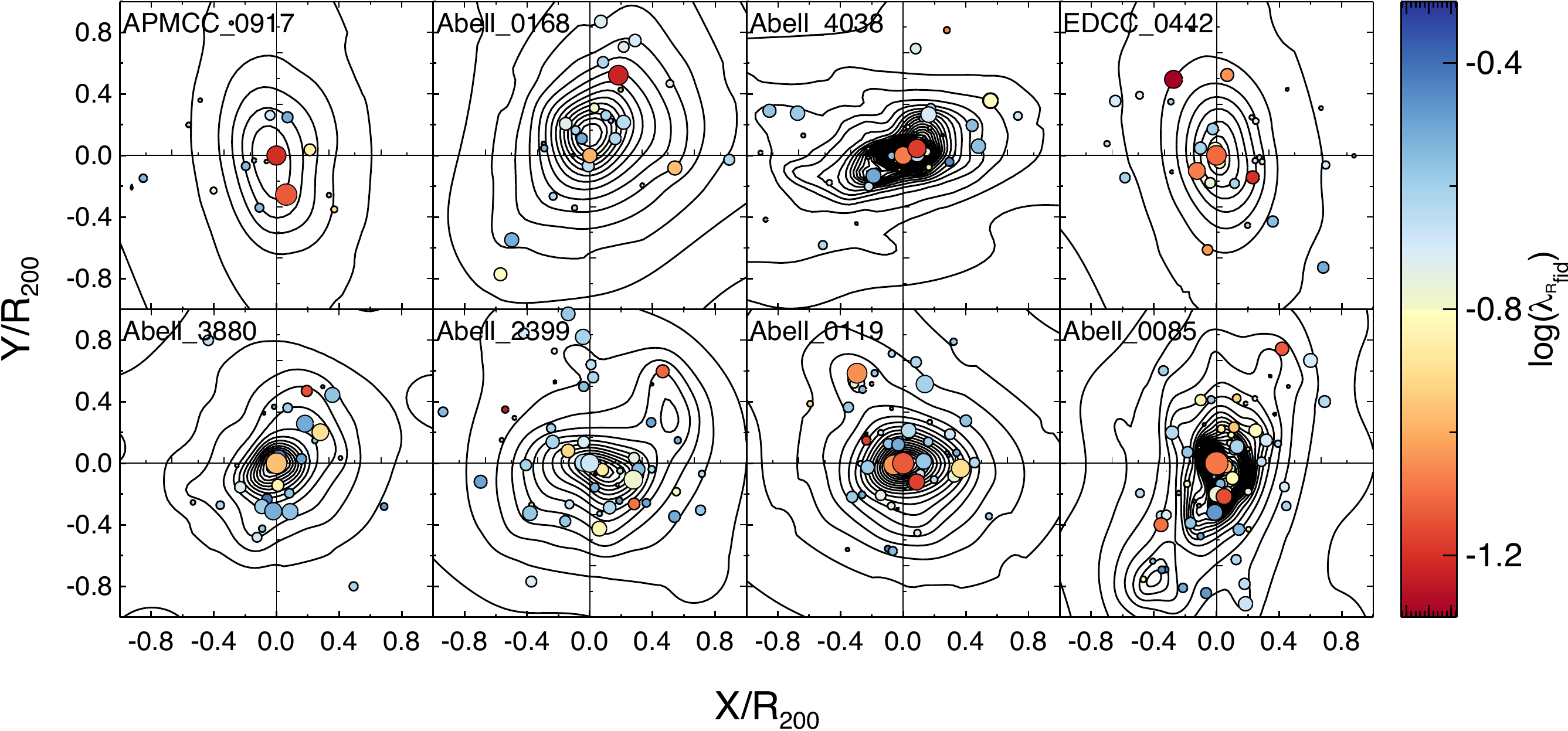}
    \caption{The spatial distribution of the observed early-type member galaxies in the 8 clusters.  The point sizes indicate stellar mass and the colors indicate $\log(\lambda_{R_{fid}})$. The black contours show galaxy isopleths that are adaptively smoothed using a varying bandwidth Gaussian kernel as described in Owers et al. (2017). The X and Y axes are in units of R$_{200}$. The slow-rotating galaxies (redder) are generally associated with the cluster centers and substructure.}
    \label{fig:spatial}
\end{figure*}

Galaxy density increases at the centers of galaxy clusters.  We test how the fraction of slow-rotating galaxies changes as a function of stacked clustercentric radius ($R_{cl}/R_{200}$) in the central panel of Figure~\ref{fig:frac3}.  We see that $F_{SR}$ increases as a function of decreasing clustercentric radius with a significance of $2.9\sigma$.  Projection effects are significant in clusters and these effects act to dilute correlations with clustercentric radius.  There is also an increase in $F_{SR}$ at $\sim0.6R_{200}$. Although this `bump' is not statistically significant we also examine the spatial distribution of the slow rotators in each cluster in Figure~\ref{fig:spatial}. While this plot will suffer from the effects of sample incompleteness as well as uncertainties in our spin parameter measurements these effects are mitigated by the higher completeness for higher stellar mass galaxies (which are more likely to be slow-rotating galaxies from the right-hand panel of Figure~\ref{fig:frac3}) and by showing the galaxies' $\lambda_{R_{fid}}$ values rather than simply whether they are fast or slow rotators.  The slow-rotating (redder) galaxies are generally located in the cluster centers ($R_{cl}<0.3R_{200}$; as indicated by the central panel of Figure~\ref{fig:frac3}) and those few that are located outside the cluster centers are generally associated with substructure in the galaxy distribution (Abell 85, 119 and 2399).  Abell 168 has a massive slow rotator at $R_{cl}>0.3R_{200}$ but does not show deviations in the galaxy distribution in Figure~\ref{fig:spatial}, however, it is a well-known merging cluster (e.g. \citealt{ulmer92}) with substructure visible in the X-rays at the position of the slow rotator \citep{fogarty14}. The EDCC 0442 cluster is an outlier to this picture with three slow-rotating galaxies located away from the cluster center ($R_{cl}\sim0.5R_{200}$). While this cluster does not have substructure visible in the smoothed galaxy distribution, in the X-ray it is a `warm'-core cluster and is likely to have only relaxed recently \citep{burns08} which could be responsible for the broader distribution of slow-rotating galaxies.

Having examined the relationship between slow rotator fraction and different measures of environment, 
we now turn to examine the relationship between slow rotator fraction and stellar mass. The right-hand panel of Figure~\ref{fig:frac3} shows that the fraction of slow-rotating galaxies increases with increasing stellar mass with a significance of $5.0\sigma$.  

We test again whether the higher SAMI observing completeness at higher stellar masses affects the $F_{SR}-M_*$ relationship. 
The mean $F_{SR,corrected}$ is shown by the dashed line in the right-hand panel of Figure~\ref{fig:frac3} and is indistinguishable from the observed relationship.

The relationship of $F_{SR}$ with mass is a more significant relationship than that seen with overdensity, suggesting that higher stellar masses could be the dominant cause of the increase in slow-rotating galaxies observed with increasing environmental density and decreasing cluster-centric radius. We explore this idea further in the next section.


\subsection{Distribution of Spin Parameter}

To explore the relationship between spin parameter, environmental density and stellar mass further we now examine the distribution of these parameters, rather than simply separating the sample into slow and fast rotators.  In order to do this we need to take into account the fact that $\lambda_{R_{fid}}$ is a projected quantity. 

\begin{figure}
	\includegraphics[width=22pc]{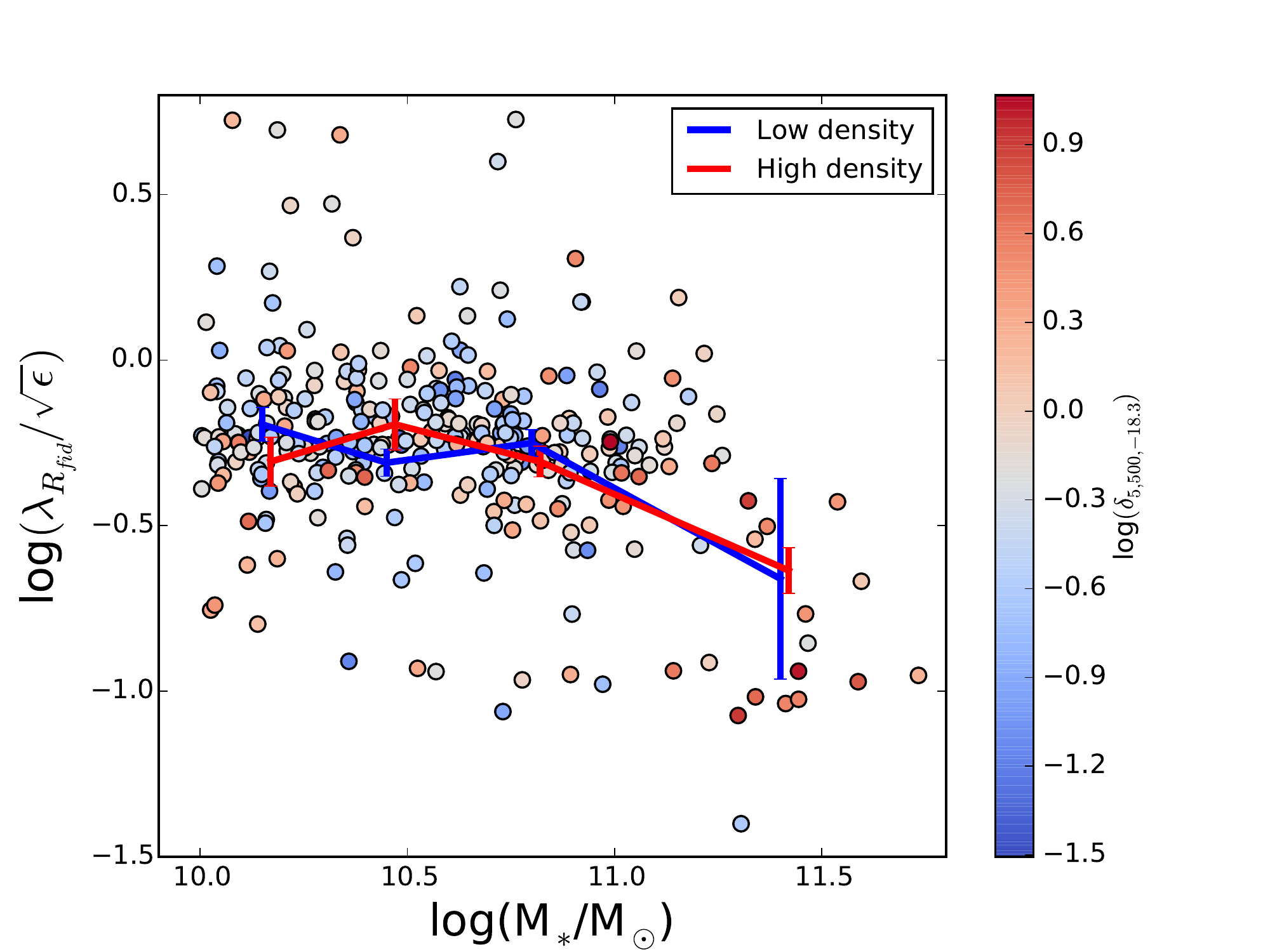}
	\includegraphics[width=22pc]{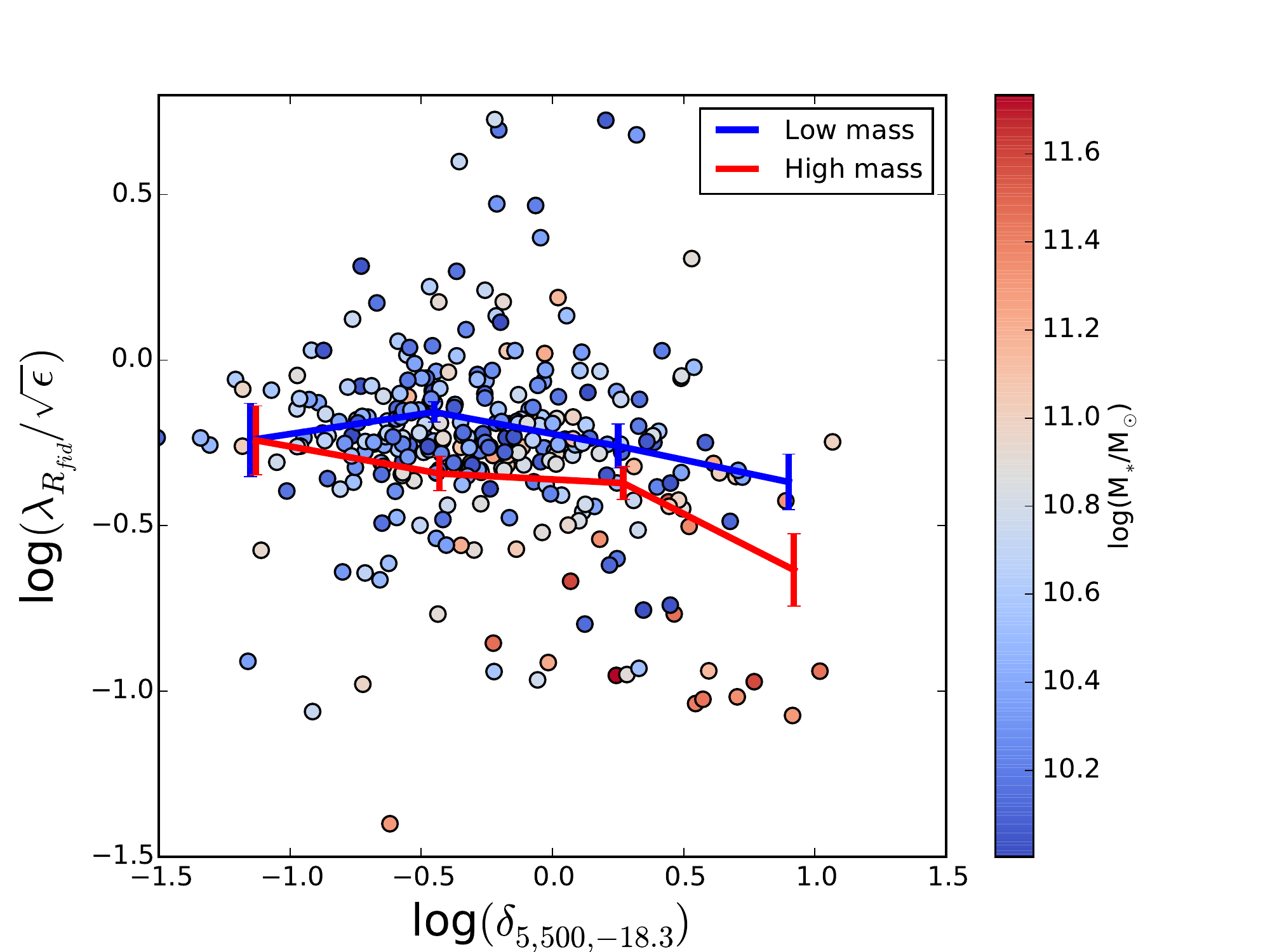}
    \caption{The upper panel shows the distribution of corrected spin parameter, $\lambda_{R_{fid}}/\sqrt\epsilon$, as a function of stellar mass, $M_*$, with colors showing environment overdensity, $\delta_{5,500,-18.3}$.  The lines show mean $\lambda_{R_{fid}}/\sqrt\epsilon$ as a function of stellar mass for the lower and upper quartiles of overdensity.  $\lambda_{R_{fid}}/\sqrt\epsilon$ does depend on stellar mass but that relationship is not significantly different between the most and least overdensities. The lower panel shows the distribution of corrected spin parameter, $\lambda_{R_{fid}}/\sqrt\epsilon$, as a function of overdensity, $\delta_{5,500,-18.3}$, with colors showing stellar mass, $M_*$.  The lines show mean $\lambda_{R_{fid}}/\sqrt\epsilon$ as a function of overdensity for the lower and upper quartiles of stellar mass.  Error bars show error on the mean calculated in bins of equal stellar mass (upper panel) and overdensity (lower panel). There is not a significant relationship between $\lambda_{R_{fid}}/\sqrt\epsilon$ and overdensity but the lower stellar mass quartile has systematically higher $\lambda_{R_{fid}}/\sqrt\epsilon$. Spin parameter depends more strongly on stellar mass than on local overdensity in clusters.}
    \label{fig:mass_lambda_env}
\end{figure}

We examine the distribution of spin parameter, $\lambda_{R_{fid}}$, applying an approximate correction for the effects of projection by dividing by ellipticity, $\sqrt\epsilon$ \citep{emsellem11}.  The upper panel of Figure~\ref{fig:mass_lambda_env} shows $\lambda_{R_{fid}}/\sqrt\epsilon$ as a function of stellar mass.  The points are colored by their environmental densities.  We also show the mean $\lambda_{R_{fid}}/\sqrt\epsilon$ as a function of stellar mass for two overdensity bins (the lowest and highest quartiles; mean overdensities $\log\delta_{low}=-0.80, \log\delta_{high}=0.35$).  
Both density bins show a relationship of decreasing $\lambda_{R_{fid}}/\sqrt\epsilon$ with increasing stellar mass, but no significant difference in that relationship as overdensity increases. We do note that the most massive galaxies, with $\log M_{*}/M_{\odot}>11.3$, have the lowest $\lambda_{R_{fid}}/\sqrt\epsilon$ and are generally in the most overdense regions ($\delta_{5,500,-18.3}>0.5$).

The lower panel of Figure~\ref{fig:mass_lambda_env} shows  $\lambda_{R_{fid}}/\sqrt\epsilon$ as a function of overdensity, $\delta_{5,500,-18.3}$ with points colored by stellar mass.  We also show the mean $\lambda_{R_{fid}}/\sqrt\epsilon$ as a function of overdensity for two stellar mass bins (the lowest and highest quartiles; mean mass \mbox{$\log M_{*,low}/M_{\odot}=10.14, \log M_{*,high}/M_{\odot}=11.08$}).  
Neither mass bin shows a strong relationship of $\lambda_{R_{fid}}/\sqrt\epsilon$ with overdensity. However, there is a systematic offset to lower $\lambda_{R_{fid}}/\sqrt\epsilon$ for the higher stellar mass sample. We also note that the most overdense regions ($\delta_{5,500,-18.3}>0.5$) are dominated by the group of massive, $\log M_{*}/M_{\odot}>11.3$, low $\lambda_{R_{fid}}/\sqrt\epsilon$ galaxies also visible in the upper panel.

A partial correlation analysis shows that the strongest relationship is between stellar mass and $\lambda_{R_{fid}}/\sqrt\epsilon$ ($R=-0.30, p=4\times10^{-8}$), with a correlation between stellar mass and environment  ($R=0.18,p=0.001$) while the relationship between $\lambda_{R_{fid}}/\sqrt\epsilon$ and surface density is not significant ($R=-0.11, p=0.04$). We conclude that the kinematic morphology--density relationship is due to the changing distribution of stellar mass with environment.

\section{Discussion}
\label{sect:discussion}
We find a total slow rotator fraction $F_{SR}=0.14\pm{0.02}$ and that this fraction does not depend significantly on host cluster mass. The lack of dependence of slow rotator fraction on global environment is consistent with previous measurements from the ATLAS$^{\rm{3D}}$ field/group sample \citep{cappellari11_2} to the massive dense cluster Abell 1689 \citep{deugenio13} that find a total $F_{SR}\sim0.15$.  

We find that $F_{SR}$ does depend on local environment, measured by overdensity, such that the fraction of slow-rotating galaxies increases as the local environment overdensity increases. The dependence of slow rotator fraction on local environment is also consistent with previous analyses \citep{cappellari11_2,deugenio13,houghton13,scott14,fogarty14}. 

We also find a strong relationship of $F_{SR}$ with galaxy stellar mass. This relationship has been observed before (e.g. \citealt{emsellem07, jimmy13, cappellari13, veale16,oliva-altamirano16}) and is not surprising given the analytic relationship between spin parameter $\lambda$, angular momentum $J$ and total mass $M$: \mbox{$\lambda=(J|E|^{1/2})/(GM^{5/2})$} (where $E$ is the total energy of the system and $G$ is the gravitational constant; e.g. \citealt{fall80,romanowsky12}). A strong relationship between specific angular momentum, $j_e$, and stellar mass has also been observed in the main SAMI Galaxy Survey by \cite{cortese16}.  

Simulations are also observing relationships between specific angular momentum and spin parameter and mass.  The analysis of specific angular momentum in the EAGLE simulation \citep{schaye15} by \cite{lagos16} also finds that it depends on stellar mass and concludes that galaxies with low $j_e$ at $z\sim0$ are a product of two pathways: galaxy mergers and early star formation quenching. Similarly, analysis of the Illustris simulation \citep{genel14} galaxies by \cite{penoyre17} finds that the slow-rotating elliptical galaxies are more massive than the fast-rotating galaxies. They also find that the slow-rotating galaxies have evolved from fast rotators since $z = 1$ as a result of mergers causing them to spin down. However, neither of these simulations include the massive cluster environments studied here. \cite{choi17} examine the evolution of the spin parameter of galaxies in cluster environments in a cosmological hydrodynamic simulation. They find that the spin evolution is mass dependent, with more massive galaxies ($\log(M_*/M_{\odot})>10.5$) experiencing more spin-down, mainly as a result of major and minor mergers. In contrast, while the spin parameter of the lowest mass galaxies ($\log(M_*/M_{\odot})<10.5$) also falls with time, this decrease is more driven by environment than by mergers. Because this mass range is at the very lowest end of our sample we cannot rule this prediction out. We also note that observations of low-mass dwarf galaxies see a strong relationship between spin parameter and environment \citep{toloba15}.

In this analysis, we have a large enough sample to disentangle the effects of local environment and stellar mass on spin parameter. When the distribution of $\lambda_{R_{fid}}$ with stellar mass is analysed together with the galaxies' local environment, we find no significant residual dependence on environment.  The lack of dependence of spin parameter on environment, once the effects of mass are removed, is in contrast to the analysis of the Fornax and Virgo clusters by \cite{scott14}. They found that even in mass-matched samples of slow and fast rotators, the slow rotators were found at higher projected environmental densities than the fast rotators. However, we note that that study was of $N\sim70$ galaxies in two low-mass clusters and our analysis of the kinematic morphology--density relationship shows that the picture in individual clusters may differ from the distribution as a whole. Our observations are consistent with more recent analyses of the classical morphology-density relationship which show that at fixed stellar mass, morphology is only weakly dependent on environment \citep{bamford09}.

Figure~\ref{fig:spatial} shows the spatial distribution of the slower and faster rotators in each of the clusters. The slow-rotating galaxies that are not within the cluster cores are generally observed to reside within substructure in those clusters.  These substructures are likely to be made up of groups that have fallen into these clusters \citep{yi13}.  We postulate that this is evidence that the kinematic morphology--density relationship is a result of mass segregation due to dynamical friction.  This evidence would suggest that slow-rotating ETGs form in a group environment which either accretes other groups over time to become a cluster, or is itself consumed to become substructure in a bigger system.  This hypothesis is consistent with the conclusions from \cite{cappellari16}.  It will be possible to test this hypothesis with the main SAMI Galaxy Survey sample (van de Sande et al, in prep) which is based on the GAMA survey of galaxies and includes a robustly-selected sample of galaxy groups \citep{robotham11}.  Examining the kinematic morphology--density relationship in the GAMA group sample will verify whether slow-rotating galaxies form in the group environment.

\section{Conclusions}
\label{sect:conclusions}
We have presented here the kinematic morphology--density relationship for a sample of 315 early-type galaxies (ETGs) in 8 galaxy clusters from the SAMI Galaxy Survey.  The 8 clusters span a halo mass range of \mbox{$14.2<\log(M_{200}/M_{\odot})<15.2$}. Cluster members were observed within $1R_{200}$ and $\pm3.5V_{gal}/\sigma_{cl}$ which covers local galaxy environments (measured by overdensity) between $-1.5<\log(\delta)\leq1.0$.  The stellar masses observed range from $10.0<\log(M_*/M_{\odot})\leq11.7$.  We classify these galaxies as fast or slow rotators depending on their spin parameter, $\lambda_{R_{fid}}$, measured from spatially-resolved stellar kinematics. We analyse the fraction of slow rotators, $F_{SR}$, as a function of local galaxy environment and stellar mass.  We also examine the distribution of $\lambda_{R_{fid}}$ as a function of both environment and stellar mass. We draw the following conclusions that are not qualitatively dependent on fiducial radius or choice of fast/slow galaxy classification:

\begin{itemize}
\item{We find a total slow rotator fraction of $F_{SR}=0.14\pm{0.02}$.} 

\item{The slow rotator fraction per cluster shows no dependence on host cluster mass in the range studied.}

\item{We find $F_{SR}$ to depend on local cluster environment such that it increases with increasing environmental overdensity, from $F_{SR}=0.14_{-0.03}^{+0.05}$ at $\log(\delta)\sim-0.9$ to \mbox{$F_{SR}=0.20_{-0.05}^{0.06}$ at $\log (\delta)\sim0.4$}, a significance of $3.4\sigma$.}

\item{$F_{SR}$ depends more strongly on stellar mass than on local cluster environment. The fraction of slow rotators increases with increasing stellar mass from \mbox{$F_{SR}=0.13_{-0.03}^{+0.06}$ at $\log (M_*/M_{\odot})\sim10.1$} to \mbox{$F_{SR}=0.41_{-0.06}^{0.07}$ at $\log (M_*/M_{\odot})\sim11.2$}, a significance of $5.0\sigma$.}

\item{Once any dependence on stellar mass is removed from the distribution of spin parameter, $\lambda_{R_{fid}}/\sqrt\epsilon$, no significant relationship with local cluster environment remains.}
\end{itemize}

We conclude that the cluster kinematic morphology--density relationship is a result of mass segregation.  We will test this hypothesis further with the broader SAMI Galaxy Survey sample (van de Sande et al., in prep).

\section*{Acknowledgements}

We thank the anonymous referee for their positive and valuable comments that have improved this paper. SB would like to thank Michele Cappellari for helpful discussions.

SB acknowledges the funding support from the Australian Research Council through a Future Fellowship (FT140101166).  JvdS is funded under Bland-Hawthorn's ARC Laureate Fellowship (FL140100278). MSO acknowledges the funding support from the Australian Research Council through a Future Fellowship Fellowship (FT140100255). NS acknowledges the support of a University of Sydney Postdoctoral Fellowship.  SMC acknowledges the support of an Australian Research Council Future Fellowship (FT100100457). This work was supported by the UK Science and Technology Facilities Council through the `Astrophysics at Oxford' grant ST/K00106X/1. RLD acknowledges travel and computer grants from Christ Church, Oxford and support from the Oxford Centre for Astrophysical Surveys which is funded by the Hintze Family Charitable Foundation. Support for AMM is provided by NASA through Hubble Fellowship grant \#HST-HF2-51377 awarded by the Space Telescope Science Institute, which is operated by the Association of Universities for Research in Astronomy, Inc., for NASA, under contract NAS5-26555.  S.K.Y. acknowledges support from the Korean National Research Foundation (2017R1A2A1A05001116) and by the Yonsei University Future Leading Research Initiative (2015-22-0064). This study was performed under the umbrella of the joint collaboration between Yonsei University Observatory and the Korean Astronomy and Space Science Institute.

The SAMI Galaxy Survey is based on observations made at the Anglo-Australian Telescope. The Sydney-AAO Multi-object Integral-field spectrograph (SAMI) was developed jointly by the University of Sydney and the Australian Astronomical Observatory, and funded by ARC grants FF0776384 (Bland-Hawthorn) and LE130100198. 

The SAMI input catalog is based on data taken from the Sloan Digital Sky Survey, the GAMA Survey and the VST ATLAS Survey. The SAMI Galaxy Survey is funded by the Australian Research Council Centre of Excellence for All-sky Astrophysics (CAASTRO), through project number CE110001020, and other participating institutions. The SAMI Galaxy Survey website is http://sami-survey.org/. 

GAMA is a joint European-Australasian project based around a spectroscopic campaign using the Anglo-Australian Telescope. The GAMA input catalogue is based on data taken from the Sloan Digital Sky Survey and the UKIRT Infrared Deep Sky Survey. Complementary imaging of the GAMA regions is being obtained by a number of independent survey programs including GALEX MIS, VST KiDS, VISTA
VIKING, WISE, Herschel-ATLAS, GMRT and ASKAP providing UV to radio coverage. GAMA is funded by the STFC (UK), the ARC (Australia), the AAO, and the participating institutions. The GAMA website is: http://www.gamasurvey.org/.

Based on data products (VST/ATLAS) from observations made with ESO Telescopes at the La Silla Paranal Observatory under program ID 177.A-3011(A,B,C).

This research has made use of the NASA/IPAC Extragalactic Database (NED), which is operated by the Jet Propulsion Laboratory, California Institute of Technology, under contract with the National Aeronautics and Space Administration. 

Funding for SDSS-III has been provided by the Alfred P. Sloan Foundation, the Participating Institutions, the National Science Foundation, and the U.S. Department of Energy Office of Science. The SDSS-III web site is http://www.sdss3.org/.



\appendix
\section{Environmental Densities}
\label{appendix:densities}

We investigated Nth nearest neighbor surface density measurements, testing the effect the choice of limits has on the environmental density we measure. We measured a suite of environmental densities, varying the Nth nearest neighbor ($N=3, 5, 10$), velocity ($V_{lim}=300, 500, 1000$ km s$^{-1}$) and absolute magnitude ($M_{lim}=-18.3, -19$ mag) limits.  We also measured the overdensity, $\delta_{N,Vlim,Mlim}=\Sigma / \bar{\Sigma}$,  dividing the density by the mean density of the early-type galaxies with $\log M_*/M_{\odot}>10$ within $1R_{200}$ of their cluster centroid. 

\begin{figure}
	\includegraphics[width=22pc]{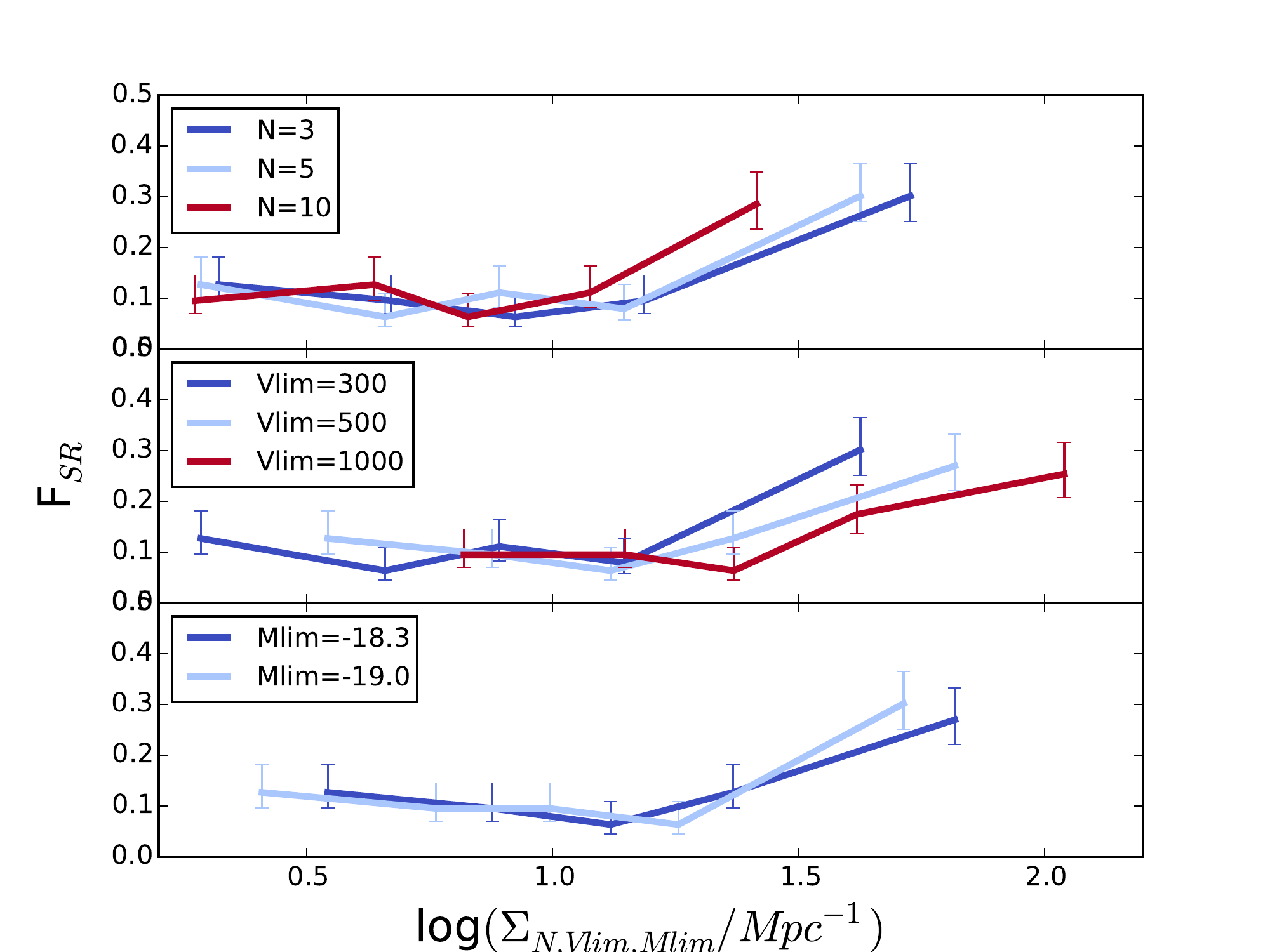}
    \caption{Slow-rotator fraction, $F_{SR}$ as a function of a suite of nearest neighbor environmental densities, $\Sigma_{N,Vlim,Mlim}$, varying the Nth nearest neighbor (upper panel; $N=3, 5, 10$; $V_{lim}=300$ km s$^{-1}$; $M_{lim}=-18.3$ mag), velocity (middle panel; $N=5$; $V_{lim}=300, 500, 1000$ km s$^{-1}$; $M_{lim}=-18.3$ mag) and absolute magnitude (lower panel; $N=5$; $V_{lim}=500$ km s$^{-1}$; $M_{lim}=-18.3, 19$ mag) limits.  The choice of limits affects the value of the environmental density measured.}
    \label{fig:env_measure}
\end{figure}
\begin{figure}
	\includegraphics[width=22pc]{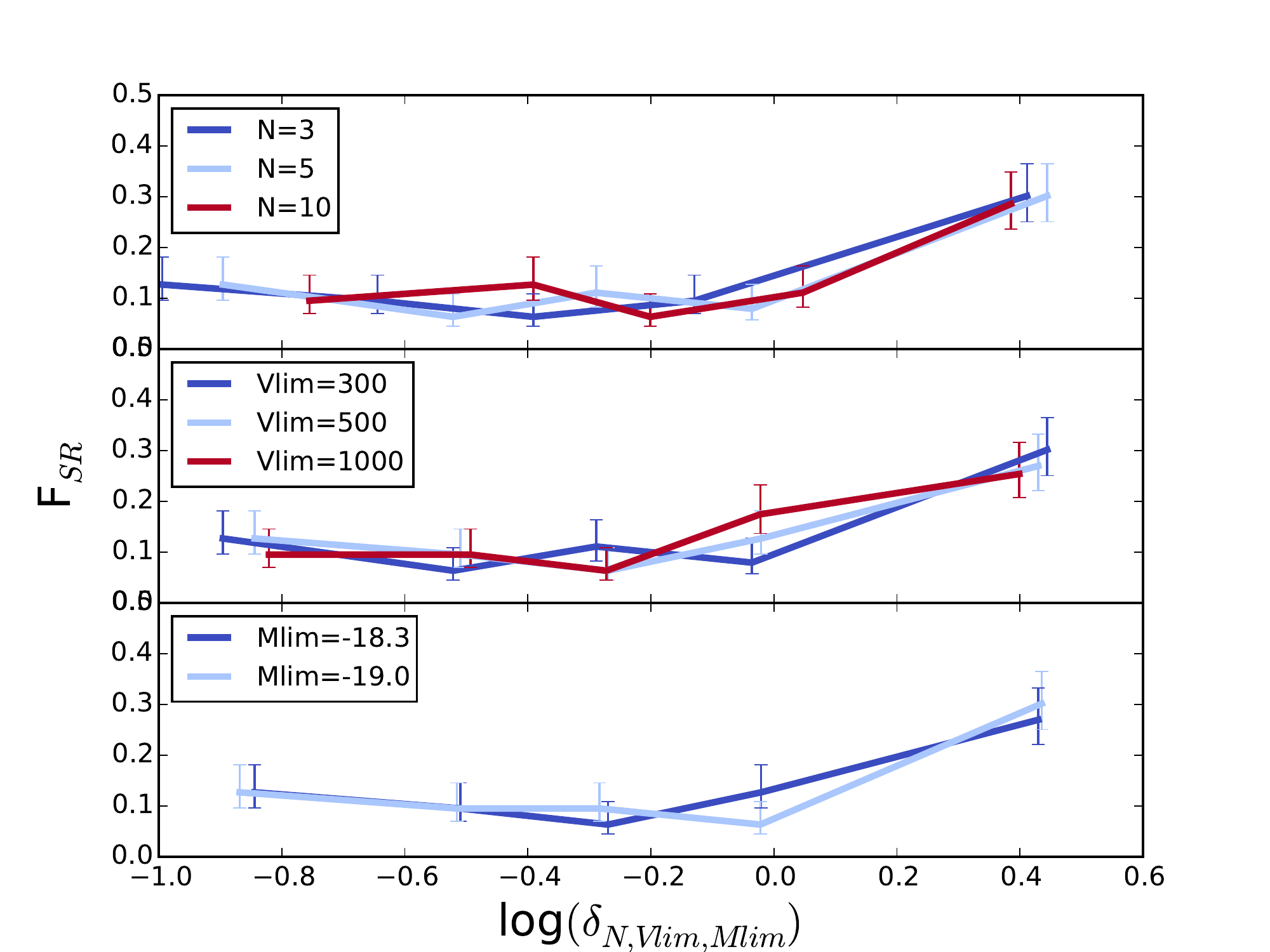}
    \caption{Slow-rotator fraction, $F_{SR}$ as a function of a suite of nearest neighbor overdensities, $\delta_{N,Vlim,Mlim}$, varying the Nth nearest neighbor (upper panel; $N=3, 5, 10$; $V_{lim}=300$ km s$^{-1}$; $M_{lim}=-18.3$ mag), velocity (middle panel; $N=5$; $V_{lim}=300, 500, 1000$ km s$^{-1}$, $M_{lim}=-18.3$ mag) and absolute magnitude (lower panel; $N=5$; $V_{lim}=500$ km s$^{-1}$; $M_{lim}=-18.3, 19$ mag) limits.  Once the nearest neighbor measurements are corrected to an overdensity they no longer depend on the limits applied.}
    \label{fig:env_over}
\end{figure}

The choice of limits affects the specific value of the environment density calculated (Figure~\ref{fig:env_measure}). This emphasises the need for caution when directly comparing the densities measured from non-homogeneous datasets that have different velocity or magnitude limits or background corrections.  However, we find the overdensities, the density divided by the mean density, to be independent of the applied limits (Figure~\ref{fig:env_over}). We therefore use the overdensity $\delta_{5,500-18.3}$ in this work.


\end{document}